\documentclass[aps,showpacs,preprintnumber,nofootinbib,amsmath,amssymb]{revtex4}

  \usepackage[latin1]{inputenc}
\usepackage{graphicx}
\usepackage{dcolumn}
\usepackage{bm}
  \usepackage{latexsym,bm,amsmath,amssymb,amsfonts}
    \usepackage{makeidx}
\newcommand{\beq}{\begin{eqnarray}}
\newcommand{\eeq}{\end{eqnarray}}
\newcommand{\x}{\vec{x}}

\begin{document}


\title{Jet energy flow at the LHC} 

\author{Yoshitaka Hatta and Takahiro Ueda }

\affiliation{ Graduate School of Pure and Applied Sciences, University
of Tsukuba, \\Tsukuba, Ibaraki 305-8571, Japan
}%


\begin{abstract}
We present a quantitative study of  energy flow away from jets by numerically solving the evolution equation derived by Banfi, Marchesini and Smye (BMS), and apply the result to two processes at the LHC: Discriminating high--$p_t$ jets originating from decays of heavy electroweak bosons from the QCD background, and the survival probability of BFKL--initiated dijet rapidity gaps. As a byproduct, we find a hidden symmetry of the BMS equation which is a remnant of conformal symmetry.
\end{abstract}

\maketitle

\section{Introduction}
 As we are entering the era of the Large Hadron Collider (LHC), the importance of jets appears more prominent than ever before. Since many of the signatures of new physics are expected to be observed via hadronic jets in disguise,
 every bit of information from jet measurements is potentially of great significance and should be carefully analyzed in light of the QCD expectations.  [See recent reviews \cite{Ellis:2007ib,Salam:2009jx} and references therein.] However,
 efforts in  this direction often encounter difficulties associated with  \emph{energy flow}---the transfer of energy or the transverse momentum away from hard jets due to the multiple emission of soft gluons. The interjet energy flow affects many observables either directly or indirectly, and tends to reduce the precision in their experimental measurements.   For instance, a small uncertainty of 1$\%$ in the jet energy can lead  to a 10 $\%$ uncertainty in the jet cross section at $p_t \sim 500$ GeV \cite{Bhatti:2005ai}.

 Normally, the soft radiation responsible for energy flow is encoded in event generators where it is no longer cleanly separable from nonperturbative effects such as the underlying event and hadronization effects. Still, one would like  to fully understand at least the perturbative part of the radiation at a quantitative level, and indeed there has been substantial progress in theory along this line over the past decade. Given such progress, it would be very interesting if one could treat energy flow as a useful tool for discovering novel phenomena at the LHC, rather than deeming it a nuisance.

 In perturbative QCD, observables related to the interjet energy flow typically involve two hard scales $E\gg E_{out} \gg \Lambda_{QCD}$ where $E$ and $E_{out}$ are the jet and interjet  energy scales, respectively. In the weak coupling expansion of such observables, logarithmically enhanced terms  of the form $(\alpha_s\ln E/E_{out})^n$ appear due to the miscancellation of real and virtual contributions. These logarithms fall into two classes which are equally important at realistic energies in collider experiments: (i) The \emph{Sudakov} logarithms arise from the direct emission of soft gluons from the primary hard partons (jets), and are thus sensitive to the antenna structure of a given process. They can be exponentiated by a sophisticated resummation procedure \cite{Kidonakis:1998nf,Oderda:1998en,Forshaw:2009fz}. (ii) The \emph{nonglobal} logarithms, first pointed out in  \cite{Dasgupta:2001sh,Dasgupta:2002bw},
are generated by soft, large--angle emissions from secondary (and ternary, etc.) gluons. These gluons multiply exponentially
 and form a  cascade in the interjet region. One then has to consider the coherent emission of soft gluons from the whole cascade whose  structure itself is determined by the previous soft radiation.
   Because of this complexity, the nonglobal logarithms do not exponentiate, but their resummation  required somewhat unconventional strategies.

  To date, two equivalent methods to resum the nonglobal logarithms have been proposed.
  Initially, Dasgupta and Salam developed a Monte Carlo simulation code to actually generate the  cascade on a computer \cite{Dasgupta:2001sh,Dasgupta:2002bw}.  This approach was  followed by several works where it was primarily used to calculate cross sections which involve rapidity gaps \cite{Appleby:2002ke,Appleby:2003sj,Delenda:2006nf}, as well as to appraise the accuracy of event generators regarding nonglobal observables \cite{Banfi:2006gy}.
 Alternatively, Banfi, Marchesini and Smye \cite{Banfi:2002hw} (BMS) have reduced the problem to solving a nonlinear integro--differential  equation. This has paved  way for the remarkable correspondence between the nonglobal logarithms and the BFKL logarithms \cite{Marchesini:2003nh,Marchesini:2004ne,Weigert:2003mm,Hatta:2008st,Avsar:2009yb}.

 In this paper we adopt the second approach and perform a detailed study of the BMS equation. We then apply the results to two processes at the LHC where energy flow plays an important role. Firstly, we suggest the possibility of  using energy flow as a discriminator of  high--$p_t$ jets initiated by highly boosted electroweak bosons from the QCD background. As recently stressed in \cite{Agashe:2008jb}, such jets may result from the decay of TeV--scale new particles, so finding methods to identify them is an urgent task at the dawn of the LHC. Secondly, we consider the perturbative survival probability of large rapidity gaps created by the BFKL Pomeron exchange. We show how to include the effect of the finite jet cone size in the gap cross section calculated in the BFKL framework. It should be said that in  these phenomenological applications, we do not intend to present a complete treatment of the problem. Rather, we shall focus on the  perturbatively calculable part of the process involving energy flow which, with more work, can be  successfully interfaced with other details of the collision event.

   In Section II we briefly review the BMS equation and show that in a certain case it admits a `bonus' symmetry which is essentially a remnant of conformal symmetry.  In Section III we solve the equation numerically and present the result in various forms that are useful for later purposes. We then devote Sections IV and V to the applications mentioned above.

\section{BMS equation and its hidden symmetry}
\subsection{The equation}

\begin{figure}[b]
\includegraphics[height=3.cm]{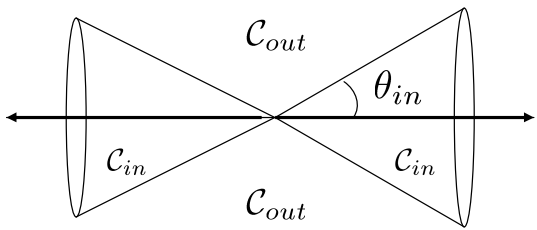}
\caption{Back--to--back jets in $e^+e^-$ annihilation.
\label{ee}}
\end{figure}

\begin{figure}
\includegraphics[height=3.5cm]{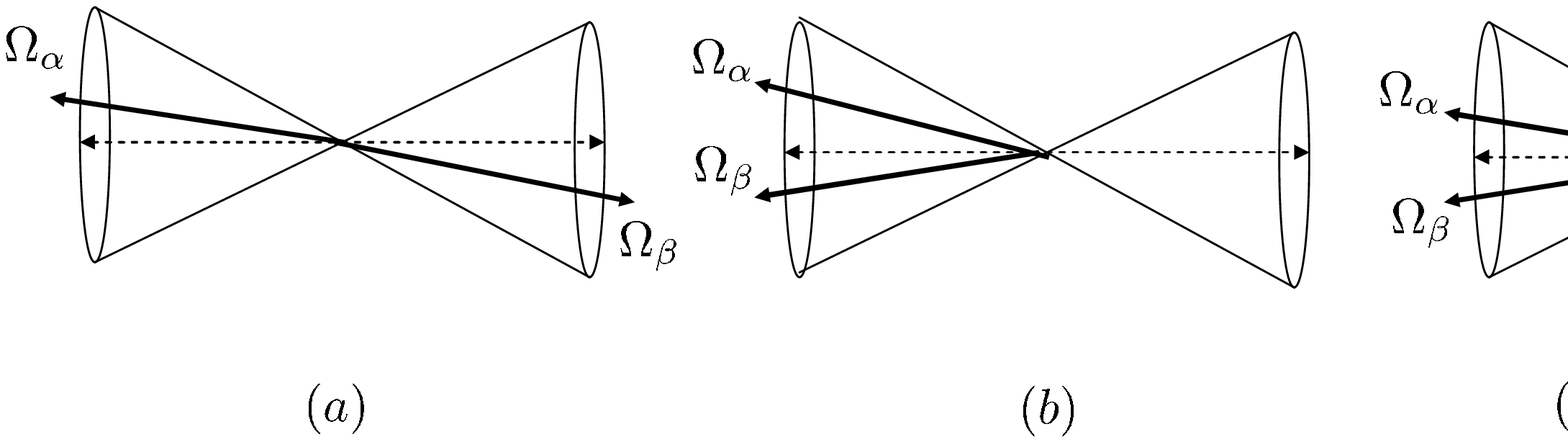}
\caption{(a) Two jets in two opposite cones. (b) Two jets in the same cone. (c) The single--cone case. In all the configurations the two jets are triggered by a color--singlet $q\bar{q}$ dipole.
\label{back}}
\end{figure}

In \cite{Banfi:2002hw}, Banfi, Marchesini and Smye (BMS) have proposed a method to quantify the efficiency of energy transfer away from hard jets summing all the single logarithmic terms in the large--$N_c$ approximation. To explain this, consider back--to--back jets with invariant mass $\sqrt{s}=2E$ produced in $e^+e^-$ annihilation (Fig.~\ref{ee}).
Define the `in' region ${\mathcal C}_{in}$ as a pair of cones with opening angle $\theta_{in}$ pointing to the thrust axes. The complementary region is denoted as  ${\mathcal C}_{out}$.
 Let $P$ be the probability that the total energy (or the total transverse momentum with respect to jets) radiated into ${\mathcal C}_{out}$  is \emph{less} than $E_{out}$
 \beq
  \sum_{i\in {\mathcal C}_{out}} E_i \le E_{out}\,.
  \eeq
It has been shown in \cite{Banfi:2002hw} that $P$ obeys a nonlinear integro--differential equation in the regime $E\gg E_{out}$, provided one  generalizes its definition to allow the two jets to point to arbitrary directions $(\Omega_\alpha,\Omega_\beta)$ inside the two cones as in Fig.~\ref{back}(a,b).
The result  is
\beq
\partial_\tau P_\tau(\Omega_\alpha,\Omega_\beta)&=& -\int_{{\mathcal C}_{out}} \frac{d^2\Omega_\gamma}{4\pi} \frac{1-\cos \theta_{\alpha\beta}}{(1-\cos \theta_{\alpha\gamma})(1-\cos \theta_{\gamma\beta})} P_\tau(\Omega_\alpha,\Omega_\beta) \nonumber \\  && \qquad  +\int_{{\mathcal C}_{in}} \frac{d^2\Omega_\gamma}{4\pi} \frac{1-\cos \theta_{\alpha\beta}}{(1-\cos \theta_{\alpha\gamma})(1-\cos \theta_{\gamma\beta})}
\Bigl(P_\tau(\Omega_\alpha,\Omega_\gamma)P_\tau(\Omega_\gamma,\Omega_\beta)-
P_\tau(\Omega_\alpha,\Omega_\beta)\Bigr)\,, \label{p}
\eeq
 where the $\theta_{in}$--dependence is implicit and the solid angle integrations are restricted as  indicated. The evolution parameter $\tau$ reads
 \beq
 \tau = \bar{\alpha}_s \ln \frac{E}{E_{out}}\,,
  \label{fix}
 \eeq
 in the fixed coupling case ($\bar{\alpha}_s \equiv \alpha_sN_c/\pi$), and
  \beq
  \tau=\frac{N_c}{\pi}\frac{1}{2b}\ln \frac{\alpha_s(E_{out})}{\alpha_s(E)}
 = \frac{N_c}{2\pi b}\ln \left(1+2b\, \alpha_s(E_{out})\ln \frac{E}{E_{out}}\right)\,, \label{run}
  \eeq
  with $b=(11N_c-2n_f)/12\pi$, in the running coupling case. The initial condition is $P_{\tau=0}=1$ for any points $(\Omega_\alpha,\Omega_\beta)$, and one also has that $P_\tau(\Omega,\Omega)=1$ for all values of $\tau$ (no radiation from a dipole with zero size).
   The important feature of the equation (\ref{p}) is that it  resums all the single--logarithmic contributions simultaneously. The Sudakov logarithms are included in the first term on the right hand side and the nonglobal logarithms in the second, nonlinear term.

 As already noted in \cite{Banfi:2002hw},  the equation can be broadly generalized to  other hard processes such as hadron--hadron collisions by  changing the definition of ${\mathcal C}_{in}$. Namely, the two cones need not be pointing back--to--back as is relevant to $e^+e^-$ annihilation in the center--of--mass frame, but their relative directions and sizes can be suitably chosen for the process of interest. [Though  this may cause complications when solving the equation numerically.] It is even allowed that ${\mathcal C}_{in}$ consists of more than two cones, or only of one cone as in Fig.~\ref{back}(c). As a matter of fact, we have found that this single--cone configuration is a particularly interesting case  both from mathematical and phenomenological points of view, and therefore it will be our main focus  in the following.

\subsection{Hidden symmetry}
\begin{figure}[h]
\includegraphics[height=9cm]{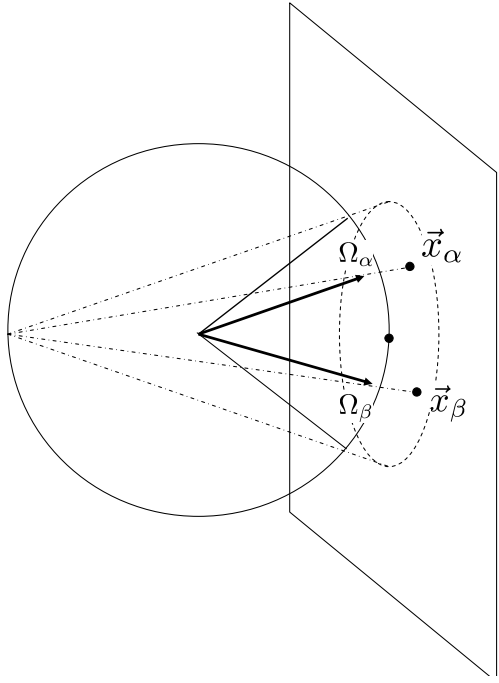}
\caption{Stereographic map between a sphere with unit diameter and a plane. The angles $(\theta,\phi)$ are measured with respect to the cone axis.
\label{map}}
\end{figure}

 Unfortunately, the nonlinear equation (\ref{p}) is too complicated to be solved analytically, though a universal feature (``geometric scaling") arises in the large--$\tau$ region for which some analytical insights can be given \cite{Banfi:2002hw}. In realistic collider experiments, there is a rather severe restriction $\tau \lesssim 1\sim 1.2\,$, and in this regime the equation has to be studied numerically. Still, here we show that in the single--cone case  there exists a hidden symmetry which puts a strong constraint on the solution.
 For this purpose, it is convenient to employ the exact correspondence \cite{Hatta:2008st,Avsar:2009yb} between the interjet soft gluon cascade and the BFKL dynamics \cite{Kuraev:1977fs,Balitsky:1978ic}.  The equation (\ref{p}) defined on a two--sphere $S^2$ with the coordinates  $\Omega=(\theta,\phi)$  can be mapped onto an equation on a two--dimensional transverse plane $\vec{x}=(x^1,x^2)$ via the stereographic projection (see, Fig.~\ref{map})
  \beq
  x^1=\tan\frac{\theta}{2}\cos \phi\,, \qquad x^2=\tan \frac{\theta}{2} \sin \phi\,.
  \eeq
 In the single--cone case, we take the origin of the polar coordinate system in the direction of the cone axis and obtain
  \beq
\partial_\tau P_\tau(\x_\alpha,\x_\beta)
= -f_{\alpha\beta}P_\tau(\x_\alpha,\x_\beta) + \int_{|\x_\gamma|<r_{in}} \frac{d^2\x_\gamma}{2\pi} \frac{(\x_\alpha-\x_\beta)^2}{(\x_\alpha-\x_\gamma)^2(\x_\gamma-\x_\beta)^2}\Bigl(P_\tau(\x_\alpha,\x_\gamma)
P_\tau(\x_\gamma,\x_\beta)
-P_\tau(\x_\alpha,\x_\beta) \Bigr)\,,
 \label{pp}
\eeq
 where
 \beq
 f_{\alpha\beta}=\int_{|\x_\gamma|>r_{in}} \frac{d^2\x_\gamma}{2\pi} \frac{(\x_\alpha-\x_\beta)^2}{(\x_\alpha-\x_\gamma)^2(\x_\gamma-\x_\beta)^2}\,. \label{integ}
 \eeq
 The solid angle integration inside/outside the cone has been mapped onto an integration inside/outside a disk with radius
 \beq
 r_{in}=\tan \frac{\theta_{in}}{2}\,.
 \eeq
Let us rescale $\vec{x}/r_{in} \to \vec{x}$ so that  $r_{in} = 1$.
If one relaxes the restriction $|\vec{x}_\gamma| \le 1$, (\ref{pp}) is identical to the Balitsky--Kovchegov (BK) equation for the dipole S--matrix $S_\tau(\vec{x}_\alpha,\vec{x}_\beta)$ which describes the gluon saturation in high energy scattering \cite{Balitsky:1995ub,Kovchegov:1999yj}.
As is well--known in that context, the integration kernel
\beq
\frac{d^2\x_\gamma}{2\pi} \frac{(\x_\alpha-\x_\beta)^2}{(\x_\alpha-\x_\gamma)^2(\x_\gamma-\x_\beta)^2}\,, \label{kernel}
\eeq
 is invariant under conformal transformation which forms the group SL(2,$\mathbb{C}$)
 \beq
 z= x^1+ix^2 \to z'=\frac{\alpha z+\beta}{\gamma z + \delta}\,, \qquad \alpha\delta-\beta\gamma=1\,.
 \eeq
 Introduction of the boundary $r_{in}=1$ breaks  conformal symmetry down to the subgroup
 SU$(1,1)\simeq$SL(2,$\mathbb{R}$). This maps the interior of the  disk  onto itself, and is defined by the following transformation
\beq
z \to z'=\frac{\alpha z+\beta}{\bar{\beta}z+\bar{\alpha}}, \qquad |\alpha|^2-|\beta|^2=1\,.
\eeq
 In the context of the BK equation, conformal symmetry is of limited use  because it is broken by the initial condition. However, the initial condition of the BMS equation does have the SU(1,1) symmetry trivially because $P_{\tau=0}(\Omega_\alpha,\Omega_\beta)=1$ for all $\Omega_{\alpha,\beta}$.
 A disk which has the isometry group  SU(1,1) is known as the Poincaré disk. The invariant measure  of the distance (the chordal distance) between two points $(\x_\alpha,\x_\beta)$ on the Poincaré disk is (see Appendix A)
 \beq
 d^2(\x_\alpha,\x_\beta)\equiv \frac{(\x_\alpha-\x_\beta)^2}{(1-\x^2_\alpha)(1-\x^2_\beta)}\,. \label{cho}
 \eeq
  Since both the equation and the initial condition are SU(1,1)--invariant, the solution $P_\tau(\x_\alpha,\x_\beta)$  must also have this symmetry which means that it is  a function only of $d^2(\x_\alpha,\x_\beta)$ (and $\tau$), or equivalently, only of the geodesic distance
 \beq
 l(\x_\alpha,\x_\beta) \equiv  \cosh^{-1}(1+2d^2(\x_\alpha,\x_\beta))=2\cosh^{-1}\sqrt{1+d^2(\x_\alpha,
 \x_\beta)}\,.
 \eeq

  In order to demonstrate the usefulness of this observation, consider the integral (\ref{integ}). This can be easily evaluated if one sets $\x_\beta=\vec{0}$. The result is, restoring $r_{in}$,
 \beq
 f_{\alpha 0}=\frac{1}{2}\ln \frac{1}{1-\frac{\x^2_\alpha}{r_{in}^2}}\,.
 \eeq
 Using the relation
  \beq
 d^2(\x_\alpha,\vec{0})=\frac{\x^2_\alpha}{r_{in}^2-\x^2_\alpha}\,,
 \eeq
  one can write
 \beq
 f_{\alpha 0}= \frac{1}{2}\ln (1+d^2(\x_\alpha,\vec{0}))\,.
 \eeq
 Since $f_{\alpha\beta}$ is a function only of $d^2(\x_\alpha,\x_\beta)$, the result with generic $\x_\beta\neq \vec{0}$ is simply given by
 \beq
 f_{\alpha\beta}=\frac{1}{2}\ln (1+d^2(\x_\alpha,\x_\beta)) &=& \frac{1}{2}\ln \left(1+\frac{r_{in}^2(\x_\alpha-\x_\beta)^2}{(r_{in}^2-\x^2_\alpha)(r_{in}^2-\x^2_\beta)} \right)
 \nonumber \\ &=&
 \frac{1}{2} \ln \left(1+\frac{\sin^2 \theta_{in}(1-\cos \theta_{\alpha\beta})}{2(\cos \theta_\alpha-\cos \theta_{in})(\cos \theta_\beta-\cos \theta_{in})} \right)\,, \label{dako}
 \eeq
   where in the second line we switched back to the original sphere problem using the stereographic projection. It requires a considerable amount of work if one tries to get the same result for $f_{\alpha\beta}$ by directly evaluating the integral (\ref{integ}) with $\x_\beta\neq 0$.

   Similarly, the full solution $P_\tau(\x_\alpha,\x_\beta)$ can be obtained from $P_\tau(\x_\alpha,\vec{0})$ which is a function only of $|\x_\alpha|/r_{in}$
   \beq
   P_\tau(\x_\alpha,\vec{0}) \equiv P_\tau\left(\frac{|\x_\alpha|}{r_{in}}\right) =P_\tau\left(\sqrt{\frac{d^2(\x_\alpha,\vec{0})}{1+d^2(\x_\alpha,\vec{0})}}\right)\,. \label{imp}
   \eeq
  $P_\tau(\x_\alpha,\x_\beta)$ for generic $\x_\beta\neq \vec{0}$ is  given by
   \beq
   P_\tau(\x_\alpha,\x_\beta)=P_\tau\left(\sqrt{\frac{d^2(\x_\alpha,\x_\beta)}{1+d^2(\x_\alpha,
   \x_\beta)}}\right)=P_\tau\left(
   \frac{|z_\alpha-z_\beta|}{
   |1-z_\alpha\bar{z}_\beta|}\right)\,. \label{wei}
     \eeq
 In practice, when solving  (\ref{pp}) (or (\ref{p})) numerically, one registers the values of $P_\tau$ for all (discretized) points  $(\vec{x}_\alpha,\vec{x}_\beta)$  at each step of iteration. This can be done straightforwardly without really caring about the constraint (\ref{wei}). The advantage of (\ref{wei}) is that a single plot of the function  $P_\tau(\vec{x}_\alpha,\vec{0})$ thus obtained tells the value of $P_\tau$ for an arbitrary point $(\vec{x}_\alpha, \vec{x}_\beta)$.


Going back to the sphere problem, (\ref{imp}) implies that
 \beq
 P_\tau(\Omega,0)=P_\tau \left(\frac{\tan \frac{\theta}{2}}{\tan \frac{\theta_{in}}{2}}\right)=P_\tau(e^{\eta_{in}-\eta})\,, \label{bst}
\eeq
 where  the pseudorapidity variable $\eta$ is defined as usual
 \beq
 \eta=\ln \cot \frac{\theta}{2}\,.
 \eeq
 (\ref{bst}) shows that, if  one of the jets is along the cone axis, $P$ becomes a function only of the relative rapidity between the other jet and the cone edge. This is of course a consequence of boost invariance. In the $\vec{x}$ coordinate,  it has a very simple meaning as the dilatation symmetry $\vec{x} \to c\vec{x}$.
 Once the cone size is fixed, the boost invariance is lost, and naively one would expect that the only symmetry of the solution $P_\tau(\Omega_\alpha,\Omega_\beta)$ would be the trivial $\phi$--rotation. Remarkably, however, a part of conformal symmetry of the soft emission probability survives, and this reduces the four degrees of freedom $(\Omega_\alpha,\Omega_\beta)$ to one.
 Note that the SU(1,1) symmetry no longer exists if there are two cones forming ${\mathcal C}_{in}$. However, we shall see that the difference in the numerical solutions with and without the backward cone turns out to be quite small.

\section{Numerical results}

In this section we present the numerical solution of the equation (\ref{p})  extending the initial result in \cite{Banfi:2002hw}.
In doing so, it is useful to factor out the Sudakov contribution
\beq
P_\tau(\Omega_\alpha,\Omega_\beta)=e^{-\tau f_{\alpha\beta}}g_\tau(\Omega_\alpha,\Omega_\beta)\,, \label{prod}
\eeq
where $f_{\alpha\beta}$ is given by   (\ref{dako}) in the single--cone case (Fig.~\ref{back}(c)), and by Eq.~(C.1) of \cite{Banfi:2002hw} in the case where two cones are pointing back--to--back (Fig.~\ref{back}(a,b)). The  effect of the non--global logarithms is then contained in $g_\tau$. At small $\tau \ll 1$, only the Sudakov contribution is important, whereas at large $\tau\gg 1$ the nonglobal contribution dominates in the sense that $e^{-\tau f}\gg g_\tau$. In the phenomenologically relevant range $\tau \sim 1$, the two contributions are comparable. Note that in the single--cone case, $f$ and $g_\tau$ are separately SU(1,1)--symmetric.

\begin{figure}[h]
\includegraphics[height=8cm]{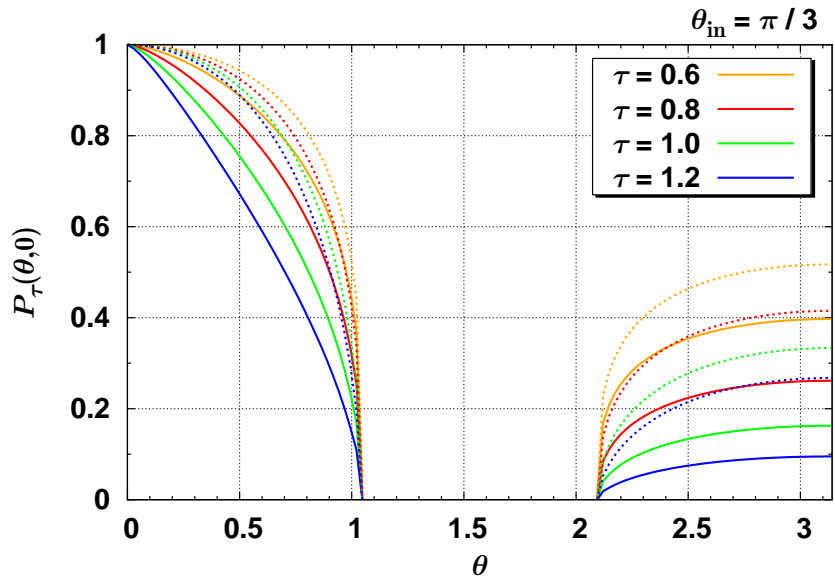}
\caption{Solid curves: the numerical solution of Eq.~(\ref{p}) in the two--cone case.  Dotted curves: the Sudakov contribution. Four curves correspond different values of $\tau$: From top to bottom, $\tau=0.6,0.8.1.0,1.2$.
\label{4}}
\end{figure}

 Let us first consider  the two--cone case. We fix $\theta_\beta=0$, $\theta_{in}=\pi/3$ and plot $P_\tau(\theta_\alpha\equiv \theta,0)$ for four values of $\tau$ between 0.6 and 1.2 in Fig.~\ref{4}. The dotted curves denote the Sudakov contribution alone.
 The result shows the expected behavior; for small $\theta$ corresponding to a dipole (dijet) with small opening angle, energy flow is suppressed due to the QCD coherence and $P$ is close to 1. On the other hand, $P$ is significantly less than 1 for dijets forming a large angle $\theta>2\pi/3$. One also sees that the nonglobal contribution becomes more important as $\tau$ gets larger.

\begin{figure}
\begin{tabular}{c}
\includegraphics[height=6cm]{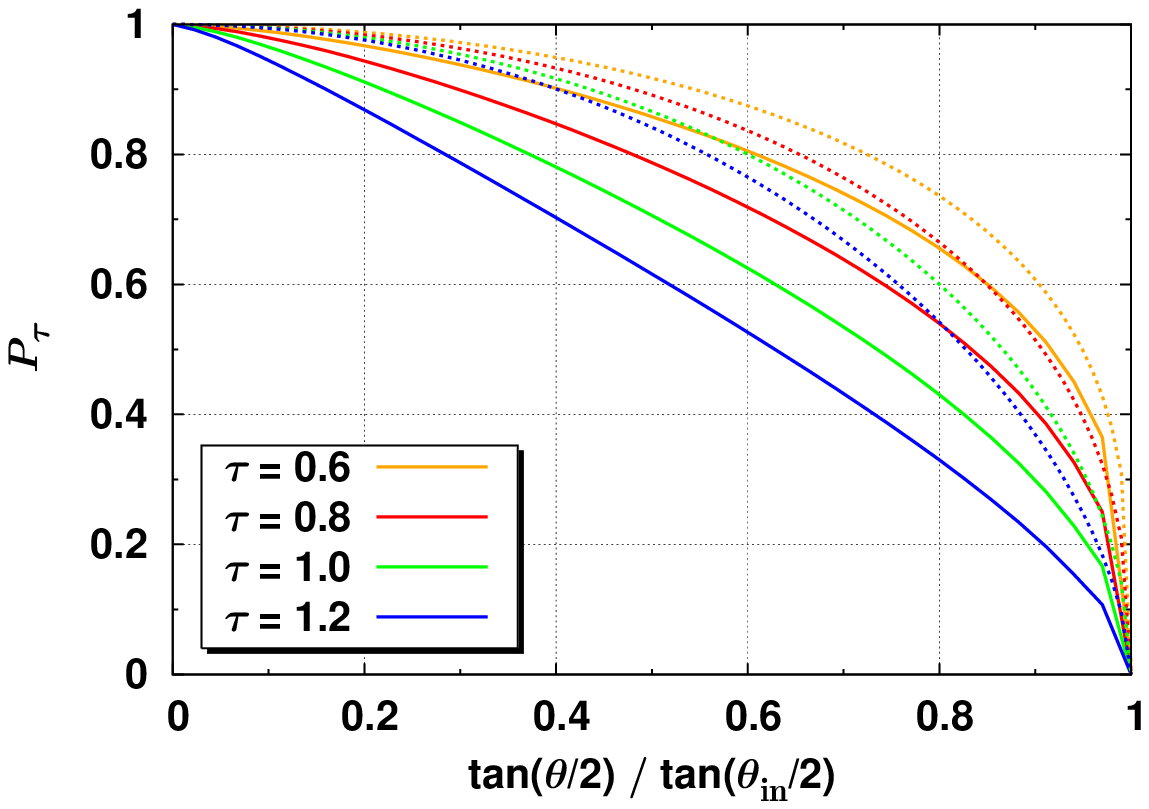}
\end{tabular}
\begin{tabular}{c}
\includegraphics[height=6cm]{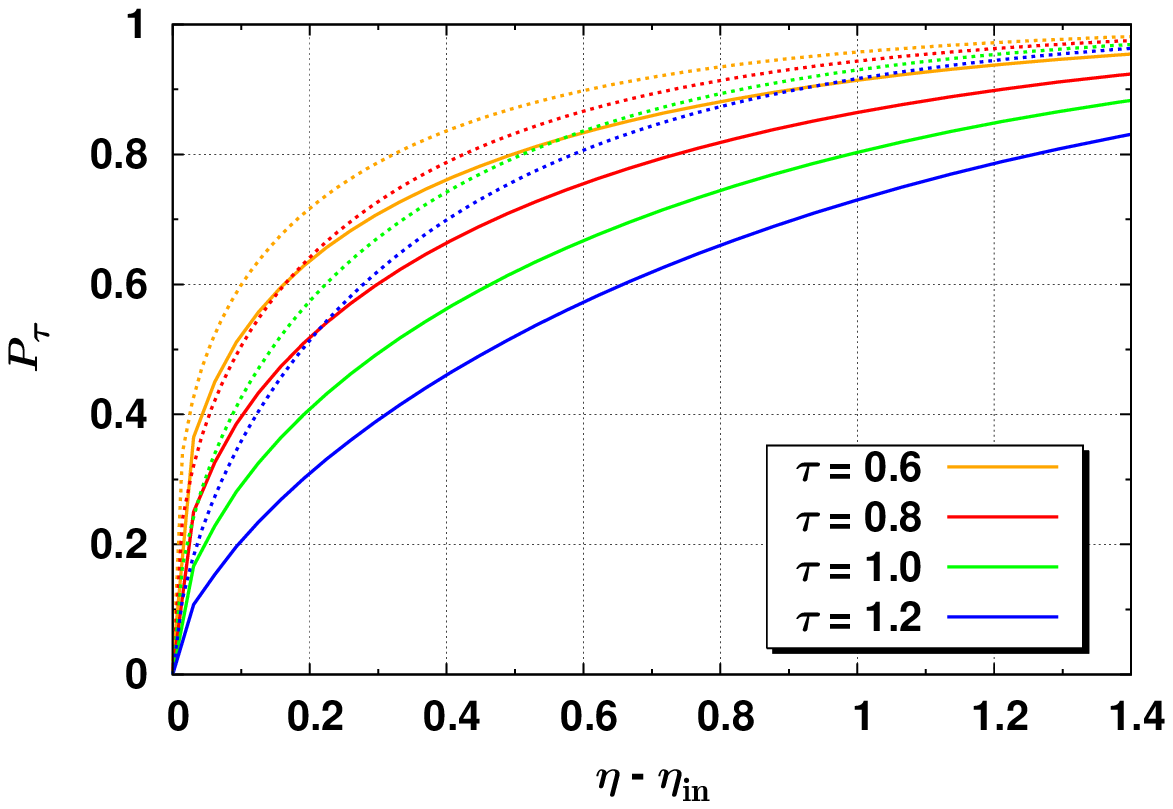}
\end{tabular}
\caption{Numerical solution of Eq.~(\ref{p}) in the single--cone case.
\label{master}}
\end{figure}

 Next, we consider the single--cone case  and show in Fig.~\ref{master} the `master function' (\ref{bst}) in the same range $0.6\le \tau \le 1.2$  as a function of $\frac{\tan \theta/2}{\tan \theta_{in}/2}$ (left) and of $\eta-\eta_{in}$ (right).
  Again the dotted curves represent the Sudakov factor which in this case reads
 \beq
e^{-\tau f}=\left(1-\frac{\tan^2\frac{\theta}{2}}{\tan^2\frac{\theta_{in}}{2}} \right)^\frac{\tau}{2}\,. \label{suda}
\eeq
In fact, numerically the result in Fig.~\ref{master} is very close to the left branch of Fig.~\ref{4}, that is, the difference between Figs.~\ref{back}(b) and \ref{back}(c) is tiny, less than 1$\%$ even for $\theta_{in}$  as large as $\pi/3$. This means that when the two jets are in the same cone the effect of the backward cone  is negligible, and practically one has the SU(1,1) symmetry to a very good approximation.

\begin{figure}[h]
\begin{tabular}{c}
\includegraphics[height=6cm]{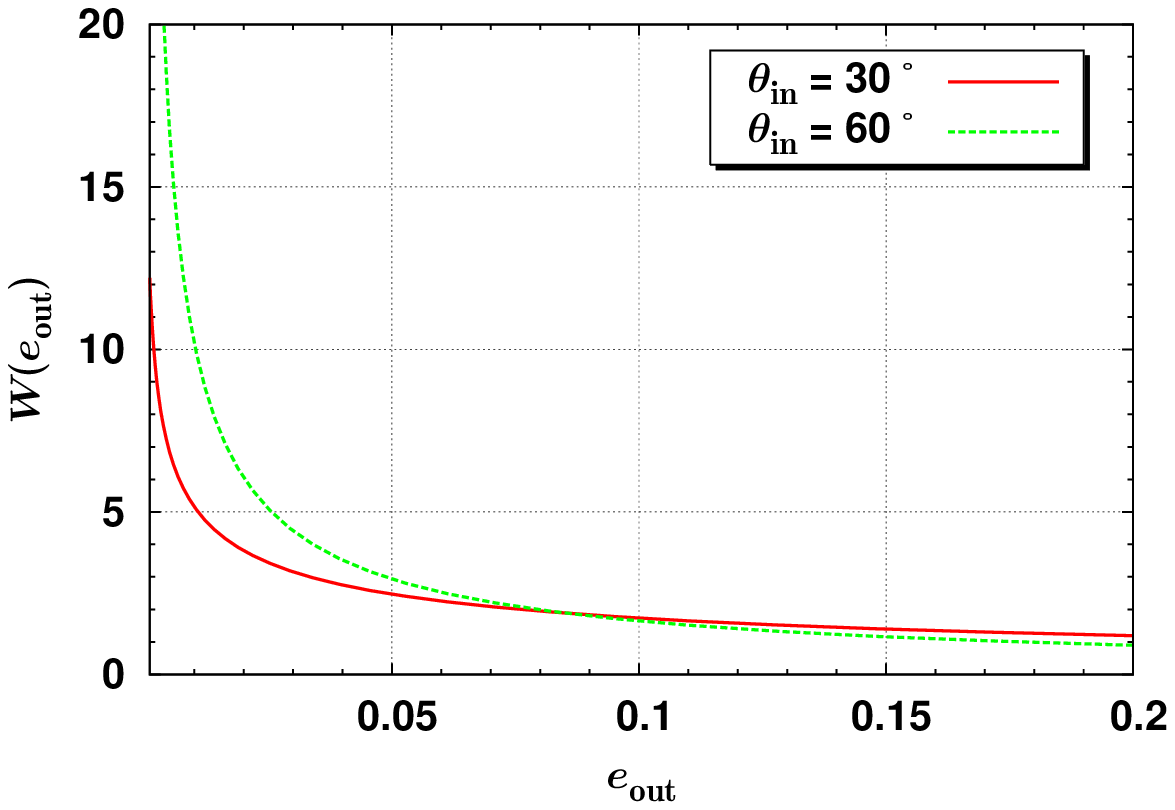}
\end{tabular}
\begin{tabular}{c}
\includegraphics[height=6cm]{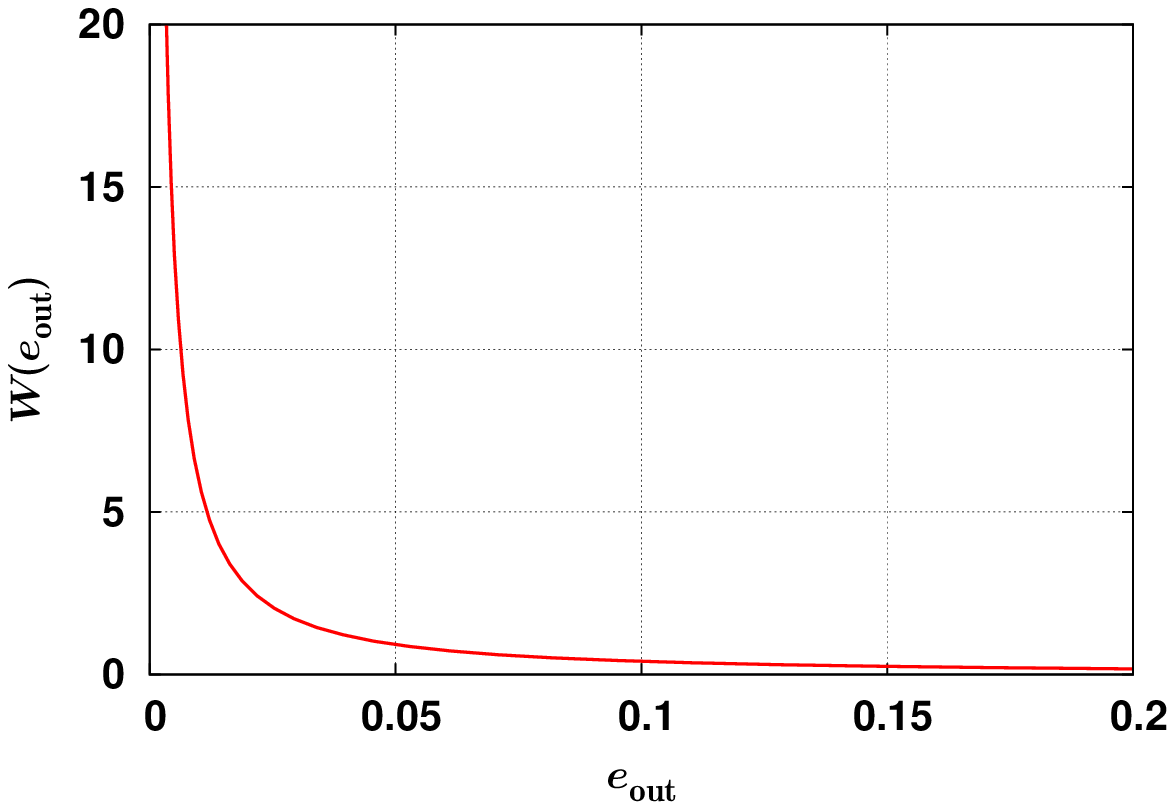}
\end{tabular}
\caption{The  probability distribution of interjet energy fraction $W(e_{out})$. Left: Two--cone case with back--to--back jets. Right: Single--cone case.
\label{5e}}
\end{figure}

Finally, we consider the probability that the interjet energy fraction
\beq
e\equiv \frac{1}{E} \sum_{i\in {\mathcal C}_{out}} E_i\,.
\eeq
is exactly $e_{out} \equiv E_{out}/E$. This may be obtained by differentiating $P_\tau$ with respect to $E_{out}$. For simplicity, we work in the fixed coupling case and denote the distribution of $e_{out}$ as $W(e_{out})$. By definition,
\beq
\int_0^{e_{out}} W(e) \,de = P_\tau\,,
\eeq
 or equivalently,
\beq
W(e_{out})=\frac{\partial}{\partial e_{out}}P_\tau = -\bar{\alpha_s}e^{\tau/\bar{\alpha}_s} \frac{\partial P_\tau}{\partial \tau}\,. \label{wwww}
\eeq
The function $W(e_{out})$ with $\bar{\alpha}_s=0.17$ is plotted in Fig.~\ref{5e}  for the back--to--back case  (left) and the single--cone case (right); the latter is obtained from (\ref{bst}) with a particular value $\tan\frac{\theta}{2}/\tan\frac{\theta_{in}}{2}= 0.46$ to be relevant later.  $W$ is sharply peaked in the small--$e_{out}$ region ($e_{out}\ll \alpha_s$) where the soft approximation is reliable, and the suppression in the `large'--$e_{out}$ region ($e_{out}\sim \alpha_s$) is stronger  in the single cone case as expected from the coherence effect. Though not shown in the figure, $W$ actually starts to decrease as  $e_{out}$ becomes extremely small ($e_{out}<10^{-5}$). This can be understood from the asymptotic large--$\tau$ analysis in \cite{Banfi:2002hw,Dasgupta:2002bw}.

Fig.~\ref{5e} shows that in the majority of  events the interjet energy flow is very small $E_{out} \ll \alpha_s E$. These typical events do not contribute to the \emph{average} of $E_{out}$ which  is dominated by rare events with $E_{out} \gtrsim \alpha_s E$.
Indeed, in the present context it is tempting to define the average of $E_{out}$ as
 \beq
\langle e_{out}\rangle=\int^{1}_0 de_{out}\, W(e_{out})\, e_{out} =  - \int^\infty_0 d\tau\, e^{-\frac{\tau}{\bar{\alpha}_s}} \frac{\partial P}{\partial \tau}
=1-\frac{1}{\bar{\alpha}_s} \int_0^\infty d\tau\, e^{-\tau/\bar{\alpha}_s} P_\tau\,,
\label{mean}
\eeq
 where in the last equality we have integrated by parts.
  Clearly, the integral is dominated by the region $\tau\lesssim \bar{\alpha}_s$ corresponding to  $E> E_{out}\gtrsim  \alpha_s E$. The soft approximation is not reliable for such large values of $E_{out}$,  one has to use instead the full splitting function to compute $\langle E_{out}\rangle$ (see, e.g., \cite{Dasgupta:2007wa}). Thus, although the probability distribution $W(e_{out})$ is meaningful at small $e_{out}\ll 1$, it is not entirely legitimate to compute the average of $e_{out}$ using (\ref{mean}). Nevertheless, the calculation is simple enough and can be done analytically since it is dominated by the Sudakov contribution. Moreover, the result may be useful to infer the qualitative feature of the differential spectrum of $E_{out}$ in the interjet region. We outline such calculations in Appendix~B.\\


Having discussed the detailed properties of the probability distribution $P_\tau$, we now turn to its phenomenological applications in collider experiments.

\section{High--$p_t$ jets from heavy electroweak bosons}
At the LHC, highly--boosted heavy electroweak bosons ($W,\,Z$, Higgs) can serve as a signal of new physics. For example, they arise as decay products of TeV--scale new particles in certain extensions of the standard model (see, \cite{Agashe:2008jb} and references therein). In order to maximize the potential of discovery, clearly it is desirable to be able to identify these events in hadronic (as well as leptonic) decay channels. Then the issue arises as to how one can experimentally distinguish, preferably  event--by--event, massive jets originating from boosted weak bosons (signal) from  QCD jets in the same mass range (background) initiated by light quarks and gluons \cite{Butterworth:2002tt,Almeida:2008yp}.  Ref.~\cite{Almeida:2008yp} looked into the substructure of jets for this purpose and showed that it has different characteristic behaviors depending on the progenitor. [For related works, see \cite{Butterworth:2008iy,Thaler:2008ju,Kaplan:2008ie,Almeida:2008tp,Ellis:2009su}.] Somewhat complementary to this, here we investigate the pattern of energy flow outside the jet cone as a possible discriminator of weak boson/QCD jets.  Our goal in this section is not to make a practical proposal that is readily useful in experiments, but rather to give the first quantitative study of the difference in energy flow between the two types of jets which will lay the foundation for future work.
 \begin{figure}
\includegraphics[height=4cm]{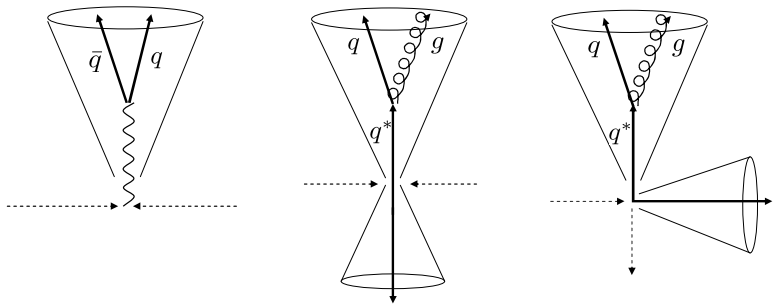}
\caption{ Left: Massive weak boson jet at midrapidity. Middle and Right: Massive  quark jet together with an accompanying jet (`color neutralizer').
\label{weak}}
\end{figure}

Fig.~\ref{weak} shows a `two--pronged' jet at midrapidity originating from  a high--$p_t$ boson (say, $Z^0$) of mass $M\ll p_t$ together with those from a quark with virtuality $M$.
The opening angle between the two primary decay particles inside the jet is bounded from below
\beq
\theta_{\alpha\beta} \gtrsim 2\frac{M}{p_t}\,, \label{typ}
\eeq
 in both cases, with the lower bound $\theta_{\alpha\beta}\sim 2M/p_t$ being the most probable configuration. Away from this peak, the two types of jets  have different particle distributions in $\theta_{\alpha\beta}$ \cite{Almeida:2008yp}.

\begin{figure}
\includegraphics[height=7cm]{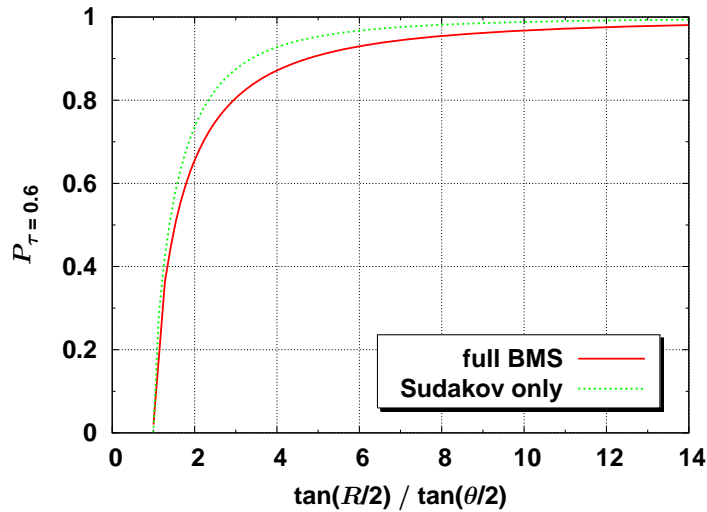}
\caption{ Red curve: The probability as a function of $\frac{\tan R/2}{\tan \theta/2}$. Green curve: Sudakov contribution (\ref{sudakov}).
\label{jet}}
\end{figure}

In the case of a weak boson jet,  radiation from the (color--singlet) $q\bar{q}$ pair is   strongly suppressed due to the QCD coherence. Let us quantify this statement by computing the probability $P_\tau$ discussed in the previous section.
 We assume that the jet is  produced around midrapidity $\eta\approx 0$ and the $q\bar{q}$ pair is symmetrically emitted so that $\Omega_\alpha=(\theta,0)$, $\Omega_\beta=(\theta,\pi)$ with $\tan \theta=M/p_t$, where the polar angles are measured with respect to the triggered jet axis. We then identify the jet cone  with ${\mathcal C}_{in}$ (i.e., the single--cone case) by taking $\theta_{in}\approx R$ where $R=\sqrt{\Delta\eta^2+\Delta\phi^2}$ is the usual jet cone radius. [Around midrapidity, $\Delta \theta \approx \Delta \eta$ for small $R=\Delta \eta$.]  Using the SU(1,1) symmetry (\ref{wei}), one can show that the configuration $\Omega_\alpha=(\theta,0)$, $\Omega_\beta=(\theta,\pi)$ is equivalent to the configuration
 \beq
 \Omega'_\alpha= (\theta',0)\,, \qquad \Omega'_\beta=(0,0)\,,
 \eeq
 where $\theta'$ satisfies
 \beq
 \frac{\tan \frac{\theta'}{2}}{\tan \frac{R}{2}}=\frac{2\frac{\tan \frac{\theta}{2}}{\tan \frac{R}{2}}}{1+\frac{\tan^2\frac{\theta}{2}}{\tan^2\frac{R}{2}}}\,. \label{eq}
 \eeq
This means that
\beq
P_{\tau}(\Omega_\alpha,\Omega_\beta)=P_\tau \left(\frac{2\frac{\tan \frac{\theta}{2}}{\tan \frac{R}{2}}}{1+\frac{\tan^2\frac{\theta}{2}}{\tan^2\frac{R}{2}}}\right)\,, \label{that}
\eeq
 namely, the probability distribution can be obtained from the master function $P_\tau(x)$ merely by the change of variables.
We take $E= p_t=1$ TeV and $E_{out}=10$ GeV with $\alpha_{s}(E_{out})\approx 0.18$ in (\ref{run}) to get $\tau\approx 0.6$. In Fig.~\ref{jet}, we have plotted $P_{\tau=0.6}$ as a function of
\beq
\frac{\tan \frac{R}{2}}{\tan \frac{\theta}{2}} \approx \frac{p_tR}{M}\,.
\eeq
The upper curve is the Sudakov contribution which in this case reads
(cf., (\ref{suda})
 and (\ref{eq}))
 \beq
\left(1-\left( \frac{2\frac{\tan \frac{\theta}{2}}{\tan \frac{R}{2}}}{1+\frac{\tan^2\frac{\theta}{2}}{\tan^2\frac{R}{2}}}\right)^2 \right)^{\frac{\tau}{2}}=\left(\frac{\frac{\tan^2 \frac{R}{2}}{\tan^2 \frac{\theta}{2}}-1}{\frac{\tan^2 \frac{R}{2}}{\tan^2 \frac{\theta}{2}}+1} \right)^\tau\,. \label{sudakov}
\eeq

Let us take, as an illustration, $M= 100$ GeV so that $p_tR/M= 10 R$.
Then Fig.~\ref{jet} shows that, if we take $R=0.4$, with 85 $\%$ probability  energy emitted outside the jet cone is less than 10 GeV. This is only 1$\%$ of the jet $p_t$ and is virtually indistinguishable from the underlying event and pileup contributions (assuming that the typical energy scale of the latter  is around this value at the LHC). The situation is similar for jets with higher $p_t$ values. Even with $p_t=4\sim 5$ TeV, $\tau$ can reach only up to $0.7$ and the function (\ref{that}) is somewhat decreased overall. However, this is more than compensated by the effect due to the decrease of the $q\bar{q}$ pair opening angle (if $M$ is kept fixed), and again one finds that with very high probability $> 90\%$, $E_{out}$ is less than 10 GeV. On the other hand, one can increase $\tau$ appreciably if $E_{out}$ is decreased, though 10 GeV might already be a bit too small value to choose in high--luminosity measurements at the LHC.
The probability distribution of $e_{out}=E_{out}/p_t$ is shown on the right hand side of Fig.~\ref{5e} in the previous section, albeit in the  fixed coupling case. As a matter of fact, this figure has been obtained for the particular $q\bar{q}$ configuration above ($p_t/M=10$, $R=0.4$).

 Next we turn to the quark jet.
  Clearly, energy flow is more efficient in this case. Due again to the QCD coherence, radiation from the $qg$ system with $\theta_{\alpha\beta}\ll 1$ is effectively that from the parent quark.  If this quark is created via the $q\bar{q}\to q\bar{q}$ hard scattering with one--gluon exchange, the compensating color is carried by the outgoing jet in the backward direction (Fig.~\ref{weak}, middle). In the $qq\to qq$ case it is carried by one of the incoming partons (Fig.~\ref{weak}, right).\footnote{There are subleading contributions in $N_c$ where the color flow assignment is interchanged between the two cases.} We thus  consider the two--cone configuration with $\theta_{in}=R$ and take   $\theta_{\alpha\beta}=\pi$ and $\pi/2$ in the two cases, respectively.  However, solving the BMS equation for the latter case is technically difficult because it involves complicated angular integrations. Rather than doing this, here we give a simple estimate of $P_\tau$ by noting that,  to lowest order in the small parameter $R$, the configuration in Fig.~\ref{weak} (right) is obtained by boosting  the back--to--back configuration with $\theta_{in}=\sqrt{2}R$  in a direction orthogonal to the jet axes with velocity $v=1/\sqrt{2}$. Of course this artificially modifies certain components of the particles' four--momenta  up to a factor $\gamma=1/\sqrt{1-v^2}=\sqrt{2}$, resulting in a small change in $\tau$ mainly via the running of the coupling. [The ratio $p_t/E_{out}$ is less affected by the boost.]  But in the present study we  ignore this change.

\begin{figure}[h]
\begin{tabular}{c}
\includegraphics[height=6cm]{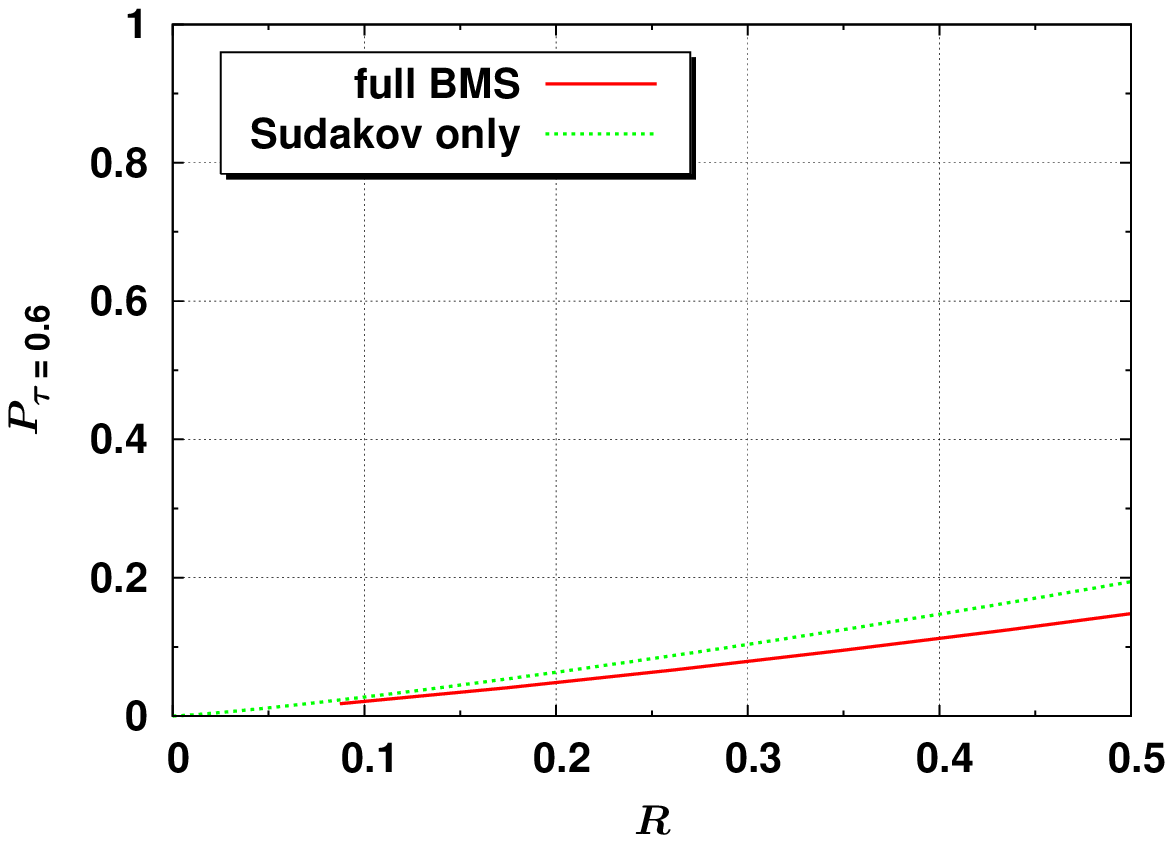}
\end{tabular}
\begin{tabular}{c}
\includegraphics[height=6cm]{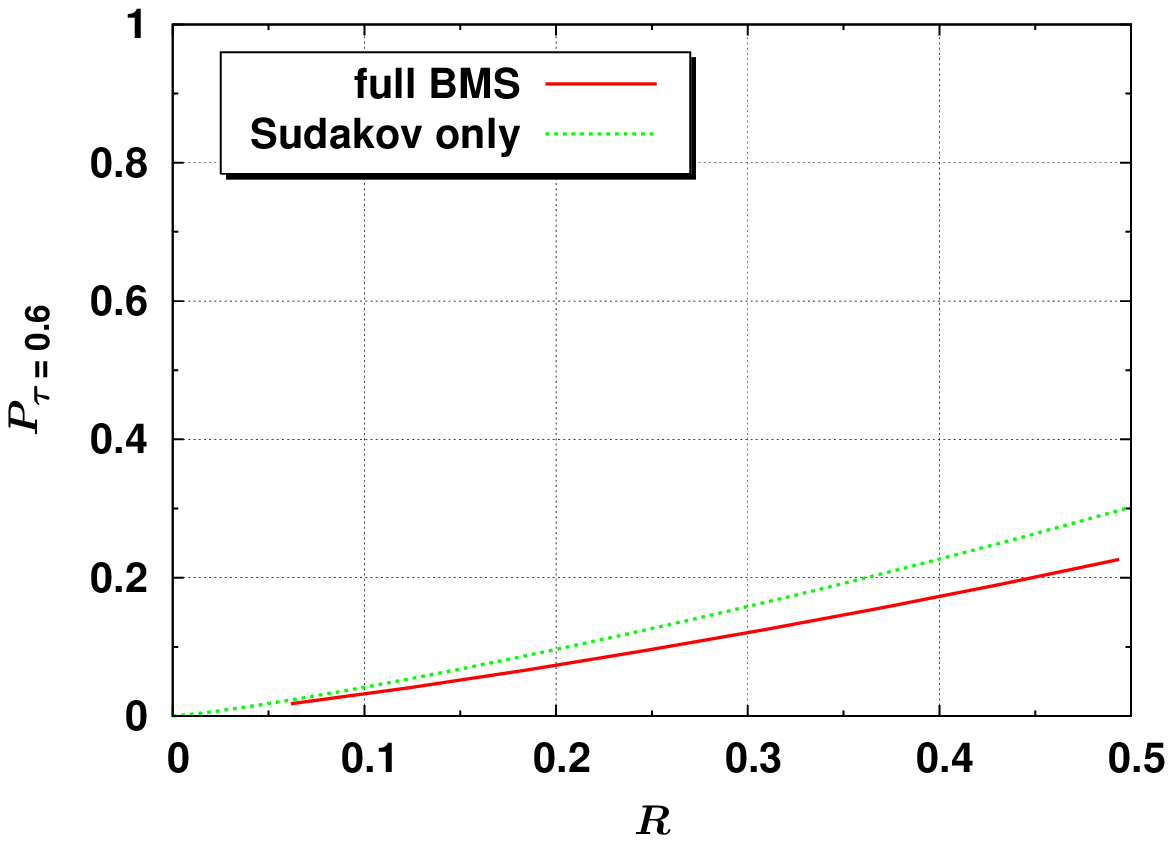}
\end{tabular}
\caption{Radiation outside the QCD jet as a function of the cone radius. The left (right) hand side corresponds to the middle (right) figure in Fig.~\ref{weak}.
\label{7}}
\end{figure}

The results are plotted in Fig.~\ref{7} as a function of $R$. In both cases, $P_\tau$ is significantly smaller, by a factor of about five,  than in the weak--jet case.  With very high probability ($80\sim 90\%$ when $R=0.4$)  the energy radiated outside the jet cone is \emph{greater} than $E_{out}=10$ GeV, though it is typically smaller than the average value $\langle E_{out}\rangle \sim \alpha_s p_t \gtrsim {\mathcal O}(100)$ GeV as can be inferred from the left hand side of Fig.~\ref{5e}.

We have thus demonstrated the  striking contrast in the amount of energy flow between jets initiated by colorless and colored particles. The five--fold difference in the probability $P_\tau$ at first looks a powerful criterion  for the event--by--event identification of jet types. [Note that the average energy is dominated by rare events, so it is not suited for the event--by--event identification.] However, in realistic experiments several other effects are expected to reduce this difference.  Most importantly, in the above we have dealt with an idealized situation where there is only one dipole containing the triggered jet, and said nothing about the presence of other dipoles in the process some of which are shown in Fig.~\ref{weak} in dotted lines. The radiation from these extra dipoles is presumably not much smaller than that from the triggered quark jet, and clearly one has to come up with a method to minimize these backgrounds. A step in the right direction would be to look at the angular region just outside the triggered jet cone  where actually most of $E_{out}$ is concentrated.\footnote{Related to this point, we note that the dipole number (not energy) distribution emitted from a boosted $q\bar{q}$ pair can be computed exactly including the nonglobal logarithms \cite{Avsar:2009yb}.} This is briefly explained in Appendix~B and will be studied further elsewhere.

\section{BFKL dijet cross section with perturbative gap survival probability}

Our second example is the production of dijets separated by a large rapidity gap $\Delta \eta$. Such `jet--gap--jet' events have been reported at the Tevatron some time ago \cite{Abbott:1998jb,Abe:1997ie} and are certainly an interesting channel to study at the LHC given the larger $p_t$ values attainable by forward jets. Two approaches to study jet--gap--jet events in perturbative QCD are commonly used:

(i) In the BFKL approach \cite{Mueller:1992pe,DelDuca:1993pq,Cox:1999dw,Enberg:2001ev,Motyka:2001zh,Chevallier:2009cu}, the exchange of the BFKL Pomeron \cite{Kuraev:1977fs,Balitsky:1978ic} in the $t$--channel naturally generates a rapidity gap, which should be the dominant process in the limit of a large gap  $\Delta \eta \gg 1$.
At the Tevatron, gap events are identified by requiring no activity (above the threshold) in the central rapidity region
 $|\eta| \le 1$.
 Yet it seems more natural to define a `gap' in the true sense of the word, to be the region between the edges of jet cones so that the actual gap size is $\Delta \eta'=\Delta \eta-2R$ where $\Delta \eta$ is the rapidity difference of dijets. This is desirable also from a theoretical point of view because in this way one can efficiently suppress the contribution from color--octet exchanges so that the signal of the BFKL Pomeron becomes more visible.  However,  jet cones filled with soft radiation are simply absent at the level of the BFKL cross section, though an event generator may be used to fix  this.

(ii) In the factorization approach \cite{Oderda:1998en,Appleby:2002ke,Appleby:2003sj,Forshaw:2005sx,Forshaw:2009fz}, one defines a gap event in terms of energy flow and starts out with the one--gluon exchange. One then dresses the amplitude with soft gluons via the renormalization group with due respect to the constraint in energy flow into the gap region. The BFKL Pomeron does not arise automatically in this approach, rather, it has to be  added by hand avoiding double counting \cite{Forshaw:2005sx}. The nonglobal logs are not included, but their effects may be estimated separately \cite{Appleby:2003sj}.

 Here we employ the  BFKL approach and discuss how to include the effect of soft radiations and of the finite jet cone size. We define gap events in terms of energy flow (as in the factorization approach) and require that the total amount of energy between the edges of jet cones is less than $E_{out}$.  Events with a perfect gap would correspond to $E_{out}=0$, but in order to use  perturbation theory one should require $E_{out}\gg \Lambda_{QCD}$. We take, somewhat optimistically, $E_{out}=1$ GeV and 2 GeV having in mind low--luminosity measurements.\footnote{Note that, as already  implicit in the previous section, in our definition $E_{out}$ includes only the perturbative radiation from the hard scatterers, that is, the contributions from the underlying event and pileups are subtracted.  These are partly taken into account by the nonperturbative survival probability ${\mathcal S}$ introduced later.}

\begin{figure}
\includegraphics[height=5cm]{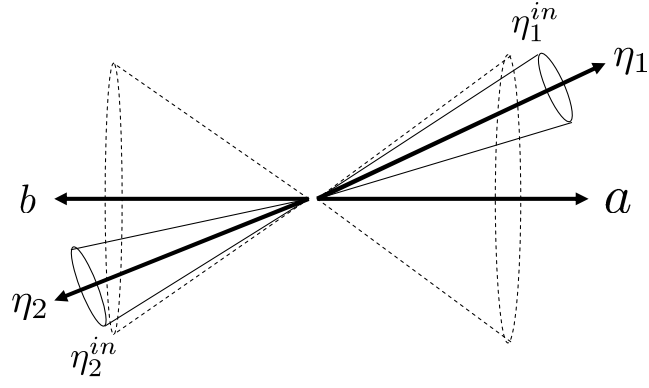}
\caption{ Radiation from two independent color dipoles.
\label{bfkl}}
\end{figure}

Specifically, consider $pp$ or $p\bar{p}$ scattering with  center--of--mass energy $s=(2E)^2$. 
We parameterize the momenta of the underlying partonic process $ab\to 12$ as (see, Fig.~\ref{bfkl})
\beq p_a&=&(x_aE,0,0,x_aE)\,, \label{pa}\\
p_b&=&(x_bE,0,0,-x_bE)\,,\\
p_1&=&(p_t\cosh \eta_1, \vec{p}_t,p_t\sinh \eta_1)\,,\\
p_2&=&(p_t\cosh \eta_2, -\vec{p}_t, p_t\sinh \eta_2)\,.
\eeq
 The relative rapidity of the two jets is $\Delta \eta=\eta_1-\eta_2$, while the average is
 $\bar{\eta}=\frac{\eta_1+\eta_2}{2}$.

The cross section in the BFKL approximation is given by ($t\approx -p_t^2$)  \cite{Mueller:1992pe}
\beq
\frac{d\sigma}{d\Delta \eta d\bar{\eta}dt} =x_ax_b\frac{d\sigma}{dx_adx_b dt} = x_a x_b f_{eff}(x_a,t) f_{eff}(x_b,t) \frac{d\sigma^{q\bar{q}\to q\bar{q}}}{dt}\,,
\label{1}
\eeq
where the effective parton distribution is ($C_F=(N_c^2-1)/2N_c$)
\beq f_{eff}(x,t)=q(x,t)+\bar{q}(x,t)+\frac{N_c^2}{C_F^2}g(x,t)\,,
\eeq
and at the leading logarithmic level,
\beq
\frac{d\sigma^{q\bar{q}\to q\bar{q}}}{dt}=\frac{4\alpha_s C_F^2}{\pi N_c^2}\frac{1}{t^2} \left(\int d\nu \frac{\nu^2}{(\nu^2+1/4)^2}e^{\bar{\alpha}_s\omega(\nu)\Delta \eta}\right)^2\,, \label{dom}
\eeq
 with $\omega(\nu)=2{\rm Re}\left[\psi(1)-\psi\left(\frac{1}{2}+i\nu\right) \right]$. Attempts to include the higher order conformal spins and the next--to--leading logarithmic effects can be found in \cite{Enberg:2001ev,Motyka:2001zh,Chevallier:2009cu}.

We now implement the survival probability of the gap against soft radiations. It is well--known that the non--forward BFKL amplitude coupled to quarks contains the Sudakov factor
  \beq
  \exp\left(-\bar{\alpha}_s\Delta \eta \ln \frac{p_t}{k_t} \right)\,,
  \eeq
   where $k_t\ll p_t$ represents the loop momentum in the soft region,
  arising from the exchange of Reggeized gluons in the $t$--channel. This limits emissions into the gap region, but apparently, the amplitude is sensitive to the infrared region. However, as explained in \cite{Bartels:1995rn} such sensitivity disappears in the Mueller--Tang formula \cite{Mueller:1992pe}. This renders the partonic cross section (\ref{dom}) truly dominated by short distance physics at the scale $\sqrt{|t|}=p_t$, thereby allowing one to treat it nearly on the same footing as, say, the one--gluon exchange contribution. Once this has been done, however, there arises additional possibility to fill the gap by soft radiation from color dipoles ($p_ap_1$) and ($p_bp_2$). [Since the BFKL exchange is color--singlet, color flows as $a\to 1$ and $b\to 2$.] This is not included in the BFKL approximation since the corresponding diagrams are not enhanced by powers of $\Delta \eta$. Still, they are enhanced by powers of $\ln p_t/E_{out}$ and can be taken into account by the survival probability $P_\tau$.

In the large $N_c$ approximation, radiations from the two dipoles ($p_ap_1$) and ($p_bp_2$) are independent. We thus modify (\ref{1}) as
 \beq
\frac{d\sigma}{d\Delta \eta d\bar{\eta} dt} = x_ax_b \tilde{f}_{eff}(x_a,t) \tilde{f}_{eff}(x_b,t) \frac{d\sigma^{q\bar{q}\to q\bar{q}}}{dt}{\mathcal S}
P_\tau(\Omega_1,\Omega_a)P_\tau(\Omega_2,\Omega_b)\,, \label{mod}
\eeq
  where
  \beq
  \tilde{f}_{eff}(x_a,t)=q(x_a,t)+\bar{q}(x_a,t)+\frac{N_c^2}{C_F^2}g(x_a,t)P_\tau(\Omega_1,\Omega_{a})\,,
  \nonumber \\
  \tilde{f}_{eff}(x_b,t)=q(x_b,t)+\bar{q}(x_b,t)+\frac{N_c^2}{C_F^2}g(x_b,t)P_\tau(\Omega_2,\Omega_b)\,.
   \label{eff} \eeq
   ${\mathcal S}$ is the nonperturbative survival probability which concerns the soft interaction among the spectators (the underlying events). It depends only on $\sqrt{s}$ and has been  estimated by several groups \cite{Gotsman:1998mm,Kaidalov:2001iz}.
     The presence of an additional factor of $P_\tau$ in front of the gluon density in (\ref{eff}) is because at large $N_c$, a gluon is represented by a double line and radiates like the square of a quark jet.

  From (\ref{bst}), one has
  \beq
  P_\tau(\Omega_1,\Omega_{a})=P_\tau(e^{\eta^{in}_1-\eta_1})=P_\tau(e^{-R})\,, \label{me}
  \eeq
 where, following our definition of a gap, we choose $\eta^{in}_1$ such that $\eta_1-\eta^{in}_1$ is always equal to $R$, see Fig.~\ref{bfkl}. Similarly, $|\eta_2-\eta^{in}_2|=R$. (\ref{me}) means that, once the cone radius is fixed, the perturbative survival probability depends only on $p_t$ (or equivalently, $\tau$) of the jets, and not on the rapidity gap $\Delta\eta$ in the present approximation.

\begin{figure}
\begin{tabular}{c}
\includegraphics[height=6cm]{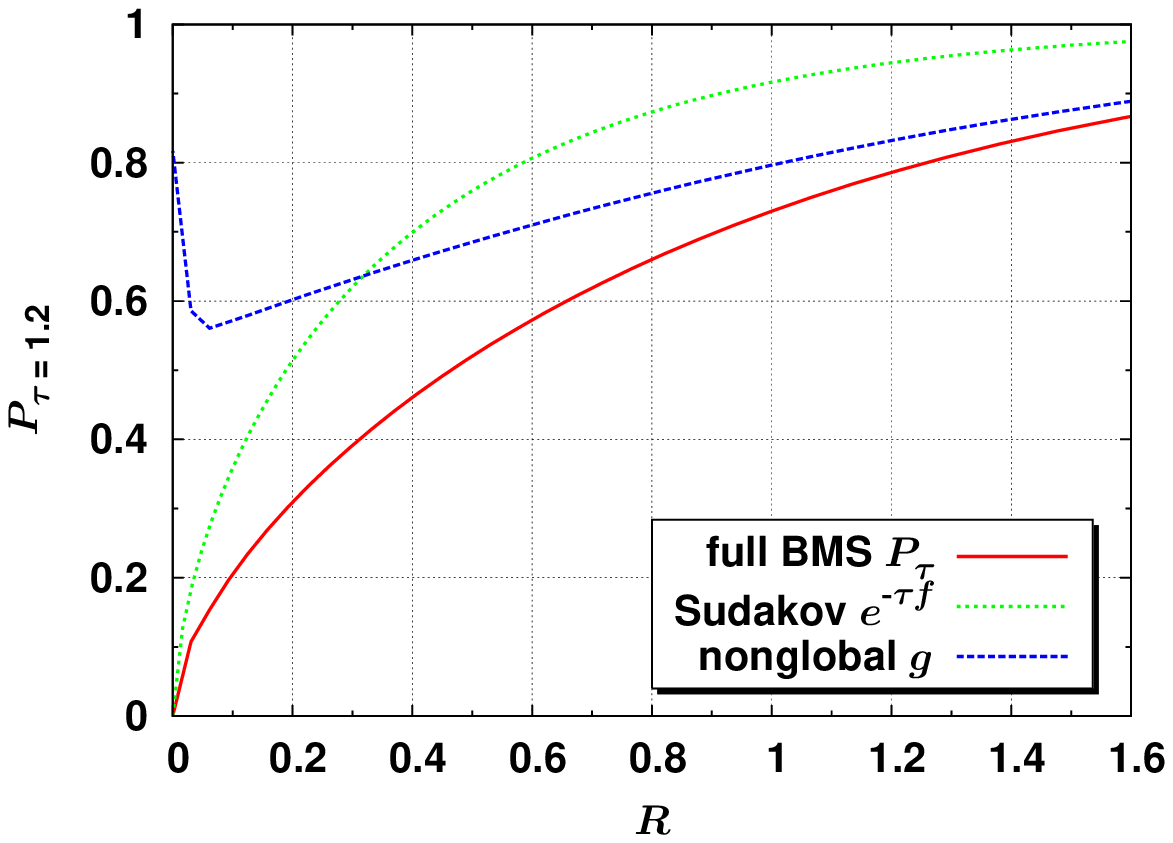}
\end{tabular}
\begin{tabular}{c}
\includegraphics[height=6cm]{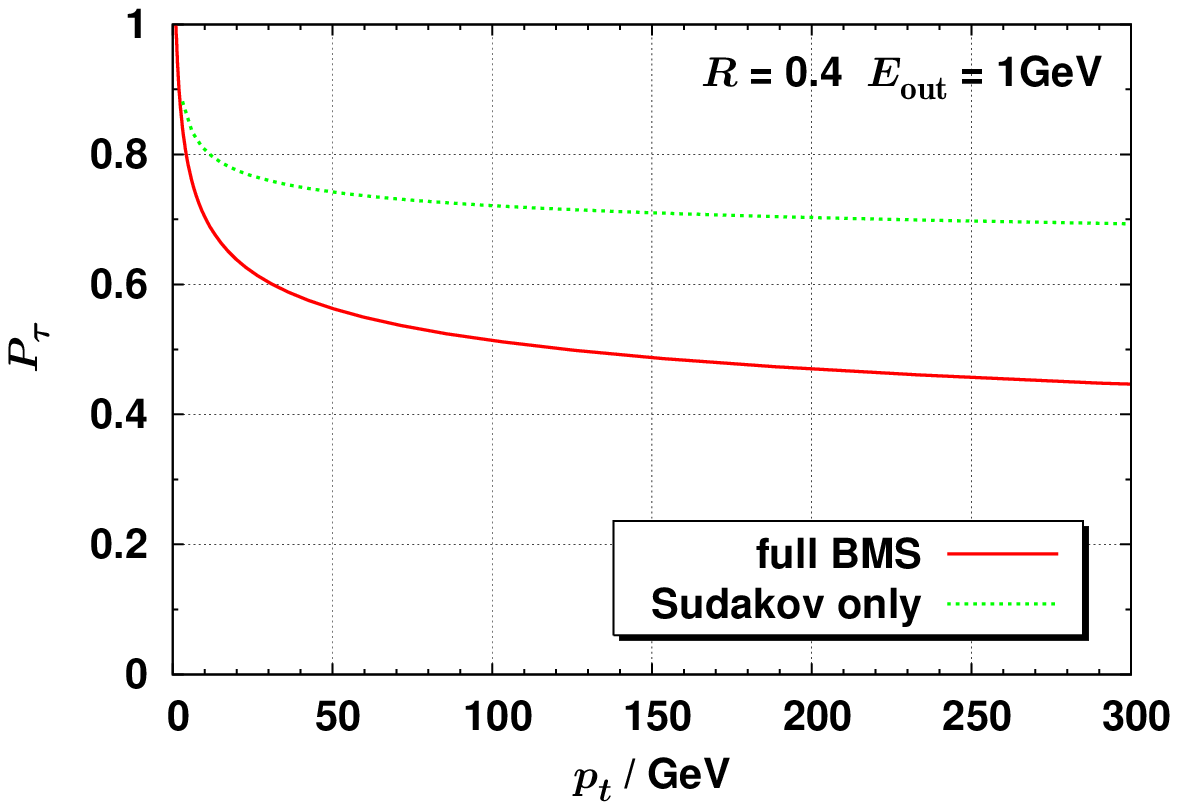}
\end{tabular}
\caption{Left: The survival probability $P_\tau$ (red line) at $\tau=1.2$ as a function of the jet cone radius $R$. The blue line is the  Sudakov factor $e^{-\tau f}$ (cf. (\ref{prod})) and the green line is the nonglobal contribution $g$. Right: The survival probability as a function of $p_t$ for fixed $R=0.4$.
\label{fig1}}
\end{figure}

\begin{figure}
\begin{tabular}{c}
\includegraphics[height=6cm]{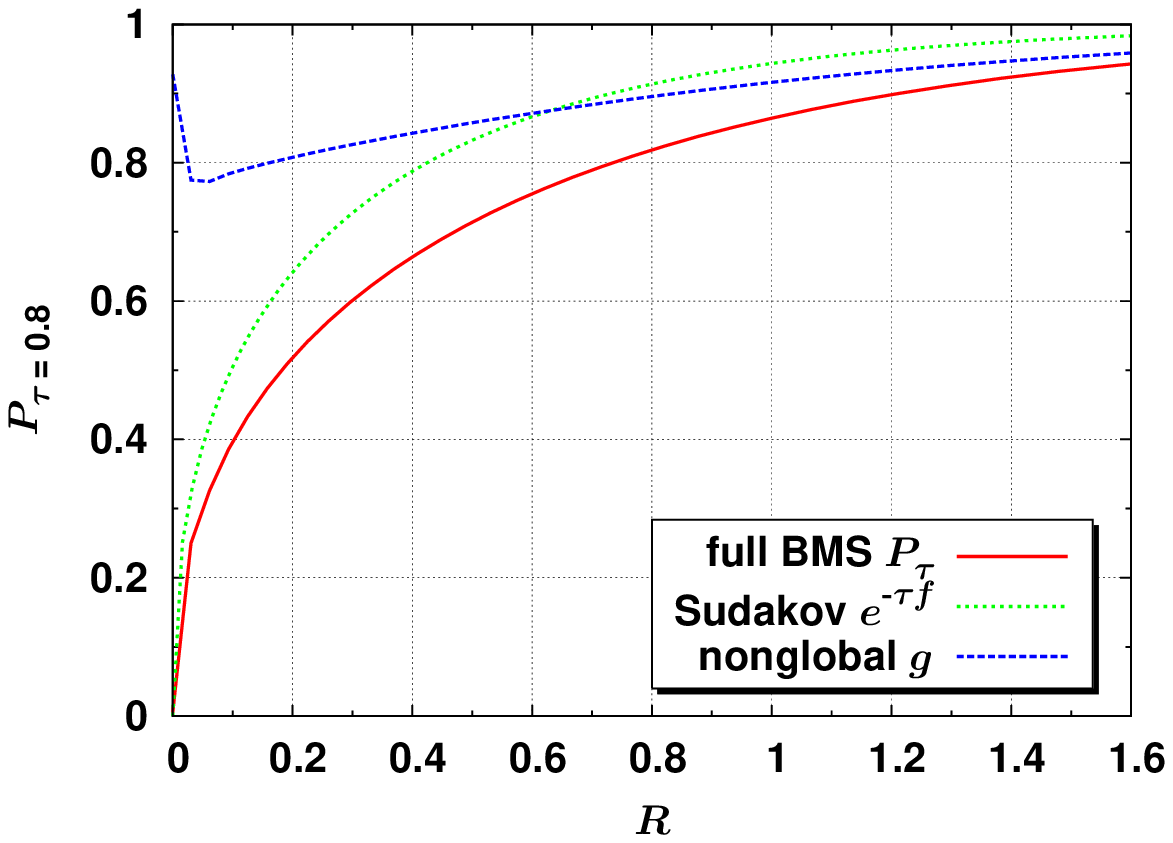}
\end{tabular}
\begin{tabular}{c}
\includegraphics[height=6cm]{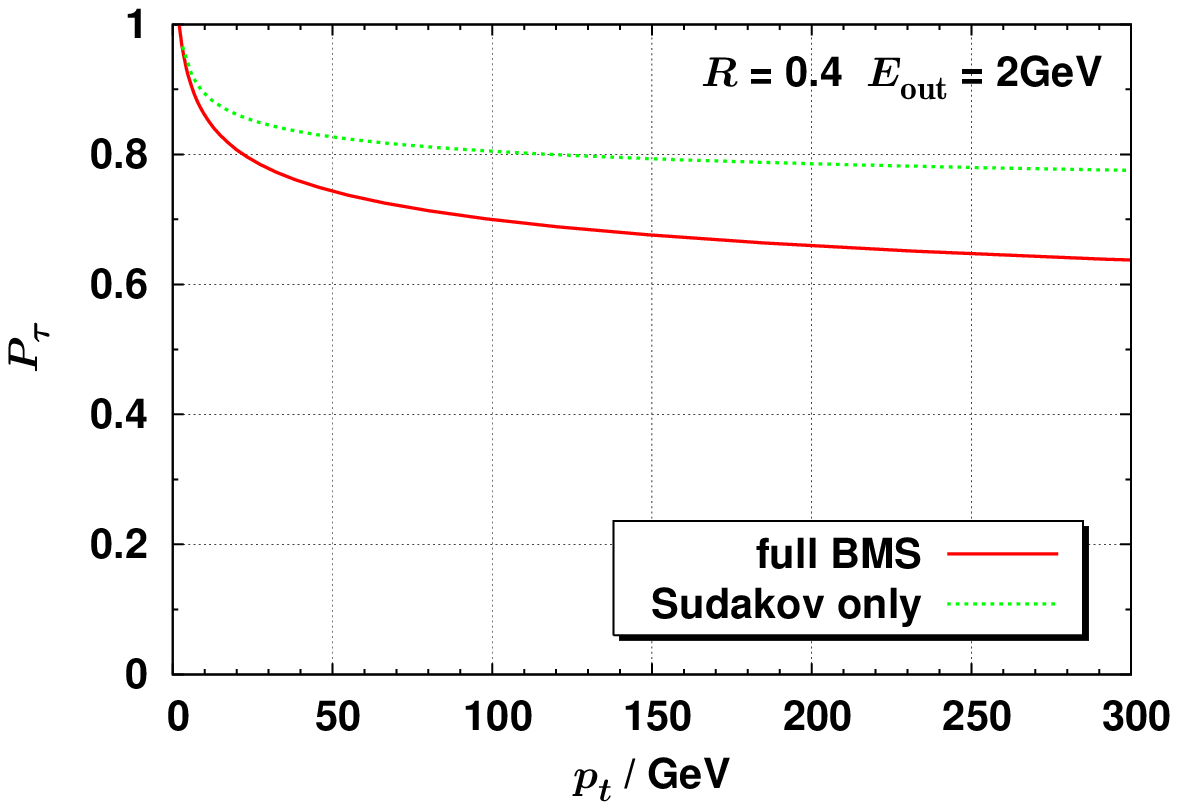}
\end{tabular}
\caption{Same as Fig.~\ref{fig1} except that $E_{out}=2$ GeV.
\label{fig2}}
\end{figure}

 On the left hand side of Fig.~\ref{fig1}, we plot  (\ref{me}) as a function of  $R$ for fixed $\tau=1.2$, corresponding to $p_t\sim 300$ GeV and $E_{out}=1$ GeV in the running coupling case with $\alpha_s(1 \mbox{GeV})\approx 0.5$.
  On the right hand side of Fig.~\ref{fig1}, we fix $R=0.4$ and plot (\ref{me}) as a function of $p_t$ using the one--loop relation between $\tau$ and $p_t$. The case with $E_{out}=2$ GeV is shown in Fig.~\ref{fig2}. We see that the results depend rather sensitively on the value of $E_{out}$ in the few--GeV region. When $E_{out}=1$ GeV the nonglobal logarithms are more important than the Sudakov logarithms.

The dependence of $P_\tau$ on jet $p_t$ is weak for $p_t > 50$ GeV and is typically $P_\tau\sim 0.5$ when $E_{out}=1$ GeV. This means that the BFKL cross section gets smaller at least by a factor $P_\tau^2\sim 0.25$. The gluon--gluon contribution is suppressed by an even smaller factor $P_\tau^4$.
 On the other hand, by requiring the gap size to grow linearly with the dijet rapidity gap $\Delta\eta'=\Delta\eta-2R$ and choosing a small $E_{out}$, one can enormously suppress the color--octet contribution. This can be readily inferred from the right branch of Fig.~\ref{4}. One sees that the typical suppression factor for the color--octet contribution is $P_\tau^2\sim 0.01$ or less when $\tau=1.2$. Thus the suppression factor $\sim 0.25$ in the color--singlet channel  appears to be a small  price to pay for a clean test of the BFKL effects. In practice, in order to get the full cross section  the above results should be combined with an event generator paying attention to the double counting of the sort discussed in \cite{Banfi:2006gy}.

\section{Conclusions}
In this paper we have presented a detailed numerical study of the BMS equation which describes the dynamics of interjet energy flow  at the single--logarithmic level in the large--$N_c$ approximation. In hard scattering processes, energy flow is often thought of as an annoying feature which, like the underlying event and the hadronization effect, blurs  otherwise clean predictions of perturbative QCD or signatures of new physics \cite{Bhatti:2005ai}. Contrary to this, our main motivation has been to demonstrate the usefulness of energy flow as a primary diagnostic tool to discover interesting physics at collider experiments. We have discussed two  examples for this purpose: Identifying high--$p_t$ jets of hadronically decaying electroweak bosons among the QCD background, and suppressing the color--octet contribution relative to the BFKL Pomeron contribution in dijet rapidity gap events. Both processes are characterized by the large ratio $p_t/E_{out}$, and we have shown that the consideration of energy flow in terms of the probability $P_\tau$ can significantly enhance the signal--to--background ratio. We also suggest that the probability $P_\tau$ is better suited to the event--by--event identification of jets than the average energy $\langle E_{out}\rangle$, since the latter does not reflect the property of the majority of events. In future it remains to integrate the results into the full collision environment to see quantitatively how the nonperturbative effects affect the perturbatively calculated  probability $P_\tau$.

\acknowledgements
We are grateful to Cyrille Marquet and Christophe Royon for many  discussions and Shinhong Kim for helpful conversations. Special thanks go to Gavin Salam for reading the manuscript and making valuable suggestions.
This work is supported, in part, by Special Coordination Funds for Promoting Science and Technology of the Ministry of Education, Culture, Sports, Science and Technology, the Japanese Government.\\

\emph{Note added. } While this paper was being finalized, a preprint \cite{Sung:2009iq} appeared which also studied energy flow as a tool to identify new particles in a different context.

\appendix
\section{Poincaré disk}
The group SU(1,1)$\cong$SO(1,2) is the isometry group of the hyperbolic space H$_2$ defined by
\beq
X_0^2-X_1^2-X_2^2=1\,.
\eeq
Via the change of the coordinates
\beq
X_0+X_2=\frac{1}{y}, \qquad x=yX_1\,,
\eeq
the space is mapped onto the upper half plane with the squared length element
\beq
ds^2=-dX_0^2+dX_1^2+dX_2^2=\frac{d\omega d\bar{\omega}}{y^2}\,,
\eeq where $\omega\equiv x+iy$ and $y\ge 0$.
 This can be further mapped onto a unit disk via the transformation
 \beq
 z=\frac{i\omega+1}{\omega+i}, \qquad ds^2=\frac{4dz d\bar{z}}{(1-|z|^2)^2}\,. \label{metr}
 \eeq
 A disk endowed with this metric is known as the Poincaré disk.
 The invariant distance between two points (the chordal distance)  on the Poincaré disk can be immediately derived from that in the H$_2$ space
 \beq
 -(X_0-X_0')^2+(X_1-X'_1)^2+(X_2-X'_2)^2=\frac{-4|\omega-\omega'|^2}{(\omega-\bar{\omega})(\omega'
 -\bar{\omega}')}
= \frac{4|z-z'|^2}{(1-|z|^2)(1-|z'|^2)}\equiv 4d^2(z,z')\,.
\eeq
 Using this, one can rewrite the kernel (\ref{kernel}) in a manifestly SU(1,1)--invariant form
 \beq
 \frac{1}{2\pi}\frac{dz_\gamma d\bar{z}_\gamma}{(1-|z_\gamma|^2)^2}
 \frac{d^2(z_\alpha,z_\beta)}{d^2(z_\alpha,z_\gamma)d^2(z_\gamma,z_\beta)}\,.
 \eeq


\section{The average energy}
In this Appendix, we estimate the rapidity (angular) distribution of $E_{out}$ in the interjet region using (\ref{mean}). As already noted, strictly speaking this goes beyond the validity range of the soft approximation. But nevertheless we wish to record the result here for  possible future applications.

\begin{figure}
\begin{tabular}{c}
\includegraphics[height=6cm]{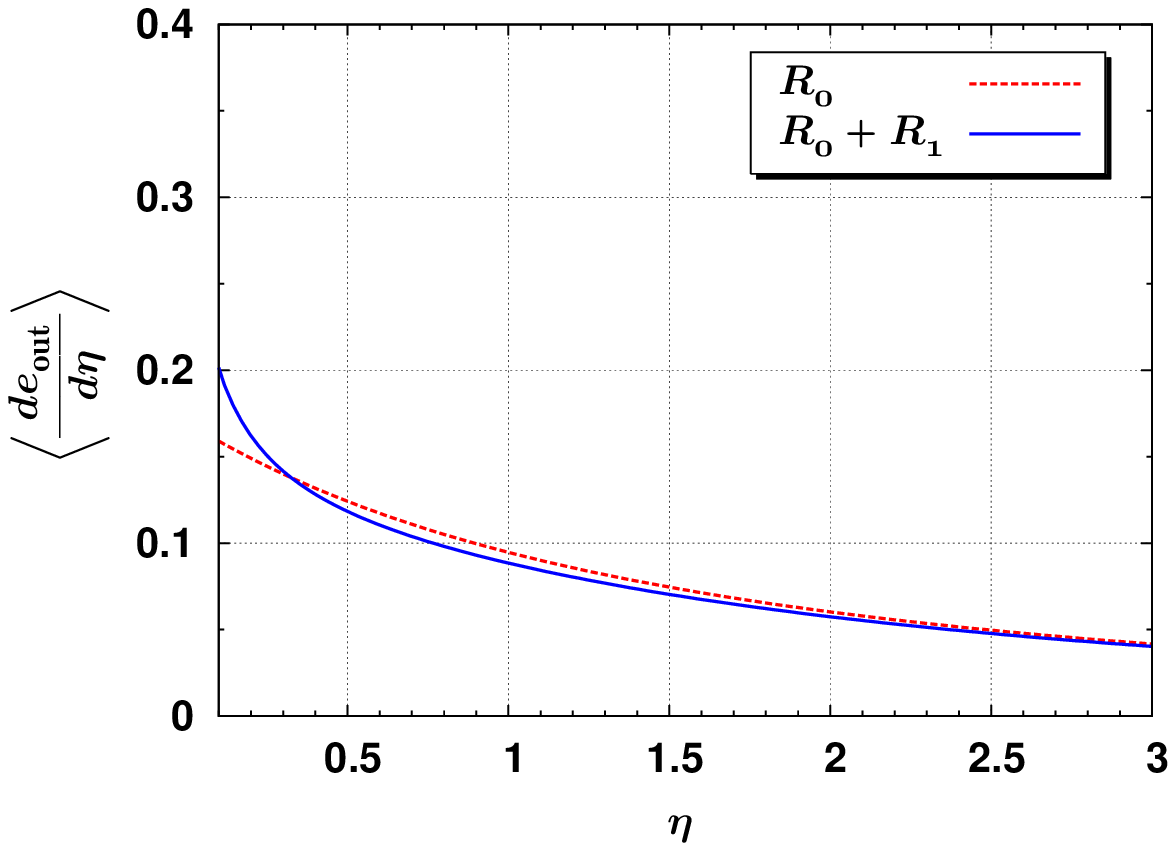}
\end{tabular}
\begin{tabular}{c}
\includegraphics[height=6cm]{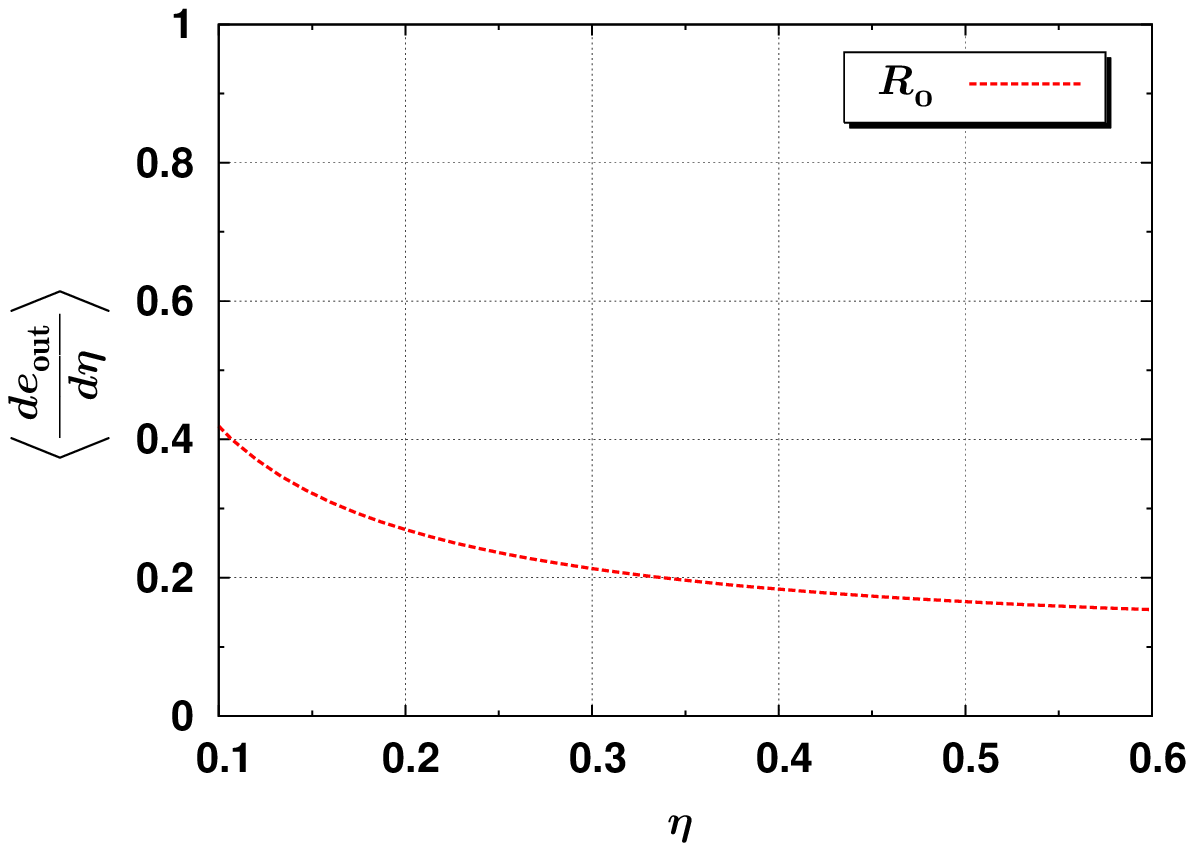}
\end{tabular}
\caption{Left: The rapidity distribution of $E_{out}$ emitted from the dipole formed by incoming partons. The region $\eta<p_t/M=0.1$ is excluded. Right: The rapidity distribution of $E_{out}$ near the quark jet.
\label{r1}}
\end{figure}

Let us begin with a dipole formed by the incoming partons as shown by dotted lines in the left and middle figures of Fig.~\ref{weak}.
Since the $\tau$--integral in (\ref{mean}) is dominated by the small--$\tau$ region, one may approximate  $P_\tau$ by its Sudakov part. Actually,  one can go one step further to include the first correction from the nonglobal logarithm \cite{Dasgupta:2002bw,Banfi:2002hw}. This gives
\beq
P_\tau(\theta_{in})\approx e^{-\tau R_0 -\tau^2 R_1}\,, \label{b1}
\eeq
where
\beq
R_0=2\ln \cot \frac{\theta_{in}}{2}=2\eta_{in}\,, \qquad
R_1=\frac{\pi^2}{12}-\frac{\mbox{Li}_2\tan^4\frac{\theta_{in}}{2}}{2} = \frac{\pi^2}{12}-\frac{\mbox{Li}_2 \,e^{-4\eta_{in}}}{2}\,.
\eeq
[$\mbox{Li}_2$ is the dilogarithm function.] Therefore,
\beq
\frac{\langle E_{out}\rangle}{p_t}=1-\frac{1}{\bar{\alpha}_s}\int_0^\infty d\tau \, e^{-\left(\frac{1}{\bar{\alpha}_s}+R_0\right)\tau-R_1\tau^2}=1-
\frac{\sqrt{\pi}}{2\bar{\alpha}_s\sqrt{R_1}}e^{\frac{\left(\frac{1}{\bar{\alpha}_s}+R_0\right)^2}{4R_1}}
\mbox{Erfc}\left(\frac{\frac{1}{\bar{\alpha}_s}+R_0}{2\sqrt{R_1}}\right)\,, \label{ave}
\eeq
where Erfc is the complementary error function.
 Let us then introduce a differential distribution
\beq
 \langle E_{out} \rangle=\int_{-\eta_{in}}^{\eta_{in}}d\eta \left\langle \frac{dE_{out}}{d\eta}\right\rangle =
  2\int_{0}^{\eta_{in}} d\eta \left\langle \frac{dE_{out}}{d\eta}\right\rangle \,,
 \eeq
such that
\beq
\left\langle \frac{dE_{out}}{d\eta}\right\rangle = \frac{1}{2}\frac{\partial \langle E_{out} \rangle}{\partial \eta_{in}}\Big\arrowvert_{\eta_{in}=\eta}\,. \label{such}
\eeq
 (\ref{such}) is plotted on the left hand side of Fig.~\ref{r1} with or without the $R_1$ term. The distribution is flat to lowest order in $\alpha_s$, but after the exponentiation it slowly varies with $\eta$. Note that $E_{out}$ in this case  should be understood as the total transverse momentum. To get the energy distribution one has to multiply the result by the factor $1/\sin \theta = \cosh \eta$.

 Next we consider the back--to--back dipole including the outgoing quark jet, Fig.~\ref{weak} (middle). This is basically the same configuration as above, but now we have to set the rapidity of the jet to be zero (rather than infinity). Then it is convenient to begin with the angular variable and  approximate $\theta_{in}\approx R$ for small $R$.
 One finds, omitting the $R_1$ term,
 \beq
 \frac{\langle E_{out}\rangle}{p_t}=\int_R^{\pi-R} \left\langle \frac{de_{out}}{d\theta}\right\rangle d\theta= \frac{2\bar{\alpha}_s \ln \cot \frac{R}{2}}{1+2\bar{\alpha}_s \ln \cot \frac{R}{2}}\approx 2\bar{\alpha}_s \ln \frac{2}{R} + {\mathcal O}(\alpha_s^2)\,. \label{all}
 \eeq
If one takes only the first term in the $\alpha_s$--expansion, there is a logarithmic divergence as $R\to 0$. This was discussed in \cite{Dasgupta:2007wa} but the coefficient here is different because we have included only the term in the splitting function which is singular in the soft limit. On the other hand, the all--order expression in (\ref{all}) is finite as $R\to 0$, and may be regarded as a partial resummation of the logarithms
$(\alpha_s\ln R)^n$. The resulting rapidity distribution $\left\langle de_{out}/d\eta\right\rangle\approx \left\langle de_{out}/d\theta\right\rangle$ is plotted on the right hand side of Fig.~\ref{r1}.
 Compared with the left hand side of Fig.~\ref{r1}, one sees that the shape is similar, but the magnitude is bigger by a factor of two in the same $\eta$ range around midrapidity.

  \begin{figure}
\includegraphics[height=5cm]{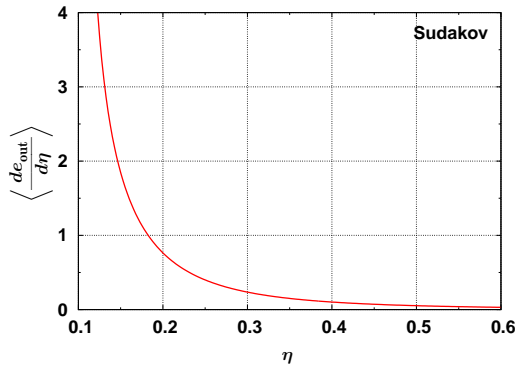}
\caption{The rapidity distribution of $E_{out}$ near a weak boson jet.
\label{final}}
\end{figure}

Finally, Fig.~\ref{final} shows the rapidity (angular) distribution of $E_{out}$ emitted from the weak boson jet with $p_t/M=10$ calculated from (\ref{sudakov}).
The distribution is very strongly peaked at small $\eta$ in striking contrast with the previous two cases. This is another way of seeing the difficulty to lose energy for a weak boson jet. See, also, \cite{Almeida:2008yp}.

\bibliography{gapref}

\begin{thebibliography}{44}
\expandafter\ifx\csname natexlab\endcsname\relax\def\natexlab#1{#1}\fi
\expandafter\ifx\csname bibnamefont\endcsname\relax
  \def\bibnamefont#1{#1}\fi
\expandafter\ifx\csname bibfnamefont\endcsname\relax
  \def\bibfnamefont#1{#1}\fi
\expandafter\ifx\csname citenamefont\endcsname\relax
  \def\citenamefont#1{#1}\fi
\expandafter\ifx\csname url\endcsname\relax
  \def\url#1{\texttt{#1}}\fi
\expandafter\ifx\csname urlprefix\endcsname\relax\def\urlprefix{URL }\fi
\providecommand{\bibinfo}[2]{#2}
\providecommand{\eprint}[2][]{\url{#2}}

\bibitem[{\citenamefont{Ellis et~al.}(2008)\citenamefont{Ellis, Huston,
  Hatakeyama, Loch, and Tonnesmann}}]{Ellis:2007ib}
\bibinfo{author}{\bibfnamefont{S.~D.} \bibnamefont{Ellis}},
  \bibinfo{author}{\bibfnamefont{J.}~\bibnamefont{Huston}},
  \bibinfo{author}{\bibfnamefont{K.}~\bibnamefont{Hatakeyama}},
  \bibinfo{author}{\bibfnamefont{P.}~\bibnamefont{Loch}}, \bibnamefont{and}
  \bibinfo{author}{\bibfnamefont{M.}~\bibnamefont{Tonnesmann}},
  \bibinfo{journal}{Prog. Part. Nucl. Phys.} \textbf{\bibinfo{volume}{60}},
  \bibinfo{pages}{484} (\bibinfo{year}{2008}), \eprint{0712.2447}.

\bibitem[{\citenamefont{Salam}(2009)}]{Salam:2009jx}
\bibinfo{author}{\bibfnamefont{G.~P.} \bibnamefont{Salam}}
  (\bibinfo{year}{2009}), \eprint{0906.1833}.

\bibitem[{\citenamefont{Bhatti et~al.}(2006)}]{Bhatti:2005ai}
\bibinfo{author}{\bibfnamefont{A.}~\bibnamefont{Bhatti}} \bibnamefont{et~al.},
  \bibinfo{journal}{Nucl. Instrum. Meth.} \textbf{\bibinfo{volume}{A566}},
  \bibinfo{pages}{375} (\bibinfo{year}{2006}), \eprint{hep-ex/0510047}.

\bibitem[{\citenamefont{Kidonakis et~al.}(1998)\citenamefont{Kidonakis, Oderda,
  and Sterman}}]{Kidonakis:1998nf}
\bibinfo{author}{\bibfnamefont{N.}~\bibnamefont{Kidonakis}},
  \bibinfo{author}{\bibfnamefont{G.}~\bibnamefont{Oderda}}, \bibnamefont{and}
  \bibinfo{author}{\bibfnamefont{G.}~\bibnamefont{Sterman}},
  \bibinfo{journal}{Nucl. Phys.} \textbf{\bibinfo{volume}{B531}},
  \bibinfo{pages}{365} (\bibinfo{year}{1998}), \eprint{hep-ph/9803241}.

\bibitem[{\citenamefont{Oderda and Sterman}(1998)}]{Oderda:1998en}
\bibinfo{author}{\bibfnamefont{G.}~\bibnamefont{Oderda}} \bibnamefont{and}
  \bibinfo{author}{\bibfnamefont{G.}~\bibnamefont{Sterman}},
  \bibinfo{journal}{Phys. Rev. Lett.} \textbf{\bibinfo{volume}{81}},
  \bibinfo{pages}{3591} (\bibinfo{year}{1998}), \eprint{hep-ph/9806530}.

\bibitem[{\citenamefont{Forshaw et~al.}(2009)\citenamefont{Forshaw, Keates, and
  Marzani}}]{Forshaw:2009fz}
\bibinfo{author}{\bibfnamefont{J.}~\bibnamefont{Forshaw}},
  \bibinfo{author}{\bibfnamefont{J.}~\bibnamefont{Keates}}, \bibnamefont{and}
  \bibinfo{author}{\bibfnamefont{S.}~\bibnamefont{Marzani}},
  \bibinfo{journal}{JHEP} \textbf{\bibinfo{volume}{07}}, \bibinfo{pages}{023}
  (\bibinfo{year}{2009}), \eprint{0905.1350}.

\bibitem[{\citenamefont{Dasgupta and Salam}(2001)}]{Dasgupta:2001sh}
\bibinfo{author}{\bibfnamefont{M.}~\bibnamefont{Dasgupta}} \bibnamefont{and}
  \bibinfo{author}{\bibfnamefont{G.~P.} \bibnamefont{Salam}},
  \bibinfo{journal}{Phys. Lett.} \textbf{\bibinfo{volume}{B512}},
  \bibinfo{pages}{323} (\bibinfo{year}{2001}), \eprint{hep-ph/0104277}.

\bibitem[{\citenamefont{Dasgupta and Salam}(2002)}]{Dasgupta:2002bw}
\bibinfo{author}{\bibfnamefont{M.}~\bibnamefont{Dasgupta}} \bibnamefont{and}
  \bibinfo{author}{\bibfnamefont{G.~P.} \bibnamefont{Salam}},
  \bibinfo{journal}{JHEP} \textbf{\bibinfo{volume}{03}}, \bibinfo{pages}{017}
  (\bibinfo{year}{2002}), \eprint{hep-ph/0203009}.

\bibitem[{\citenamefont{Appleby and Seymour}(2002)}]{Appleby:2002ke}
\bibinfo{author}{\bibfnamefont{R.~B.} \bibnamefont{Appleby}} \bibnamefont{and}
  \bibinfo{author}{\bibfnamefont{M.~H.} \bibnamefont{Seymour}},
  \bibinfo{journal}{JHEP} \textbf{\bibinfo{volume}{12}}, \bibinfo{pages}{063}
  (\bibinfo{year}{2002}), \eprint{hep-ph/0211426}.

\bibitem[{\citenamefont{Appleby and Seymour}(2003)}]{Appleby:2003sj}
\bibinfo{author}{\bibfnamefont{R.~B.} \bibnamefont{Appleby}} \bibnamefont{and}
  \bibinfo{author}{\bibfnamefont{M.~H.} \bibnamefont{Seymour}},
  \bibinfo{journal}{JHEP} \textbf{\bibinfo{volume}{09}}, \bibinfo{pages}{056}
  (\bibinfo{year}{2003}), \eprint{hep-ph/0308086}.

\bibitem[{\citenamefont{Delenda et~al.}(2006)\citenamefont{Delenda, Appleby,
  Dasgupta, and Banfi}}]{Delenda:2006nf}
\bibinfo{author}{\bibfnamefont{Y.}~\bibnamefont{Delenda}},
  \bibinfo{author}{\bibfnamefont{R.}~\bibnamefont{Appleby}},
  \bibinfo{author}{\bibfnamefont{M.}~\bibnamefont{Dasgupta}}, \bibnamefont{and}
  \bibinfo{author}{\bibfnamefont{A.}~\bibnamefont{Banfi}},
  \bibinfo{journal}{JHEP} \textbf{\bibinfo{volume}{12}}, \bibinfo{pages}{044}
  (\bibinfo{year}{2006}), \eprint{hep-ph/0610242}.

\bibitem[{\citenamefont{Banfi et~al.}(2007)\citenamefont{Banfi, Corcella, and
  Dasgupta}}]{Banfi:2006gy}
\bibinfo{author}{\bibfnamefont{A.}~\bibnamefont{Banfi}},
  \bibinfo{author}{\bibfnamefont{G.}~\bibnamefont{Corcella}}, \bibnamefont{and}
  \bibinfo{author}{\bibfnamefont{M.}~\bibnamefont{Dasgupta}},
  \bibinfo{journal}{JHEP} \textbf{\bibinfo{volume}{03}}, \bibinfo{pages}{050}
  (\bibinfo{year}{2007}), \eprint{hep-ph/0612282}.

\bibitem[{\citenamefont{Banfi et~al.}(2002)\citenamefont{Banfi, Marchesini, and
  Smye}}]{Banfi:2002hw}
\bibinfo{author}{\bibfnamefont{A.}~\bibnamefont{Banfi}},
  \bibinfo{author}{\bibfnamefont{G.}~\bibnamefont{Marchesini}},
  \bibnamefont{and} \bibinfo{author}{\bibfnamefont{G.}~\bibnamefont{Smye}},
  \bibinfo{journal}{JHEP} \textbf{\bibinfo{volume}{08}}, \bibinfo{pages}{006}
  (\bibinfo{year}{2002}), \eprint{hep-ph/0206076}.

\bibitem[{\citenamefont{Marchesini and Mueller}(2003)}]{Marchesini:2003nh}
\bibinfo{author}{\bibfnamefont{G.}~\bibnamefont{Marchesini}} \bibnamefont{and}
  \bibinfo{author}{\bibfnamefont{A.~H.} \bibnamefont{Mueller}},
  \bibinfo{journal}{Phys. Lett.} \textbf{\bibinfo{volume}{B575}},
  \bibinfo{pages}{37} (\bibinfo{year}{2003}), \eprint{hep-ph/0308284}.

\bibitem[{\citenamefont{Marchesini and Onofri}(2004)}]{Marchesini:2004ne}
\bibinfo{author}{\bibfnamefont{G.}~\bibnamefont{Marchesini}} \bibnamefont{and}
  \bibinfo{author}{\bibfnamefont{E.}~\bibnamefont{Onofri}},
  \bibinfo{journal}{JHEP} \textbf{\bibinfo{volume}{07}}, \bibinfo{pages}{031}
  (\bibinfo{year}{2004}), \eprint{hep-ph/0404242}.

\bibitem[{\citenamefont{Weigert}(2004)}]{Weigert:2003mm}
\bibinfo{author}{\bibfnamefont{H.}~\bibnamefont{Weigert}},
  \bibinfo{journal}{Nucl. Phys.} \textbf{\bibinfo{volume}{B685}},
  \bibinfo{pages}{321} (\bibinfo{year}{2004}), \eprint{hep-ph/0312050}.

\bibitem[{\citenamefont{Hatta}(2008)}]{Hatta:2008st}
\bibinfo{author}{\bibfnamefont{Y.}~\bibnamefont{Hatta}},
  \bibinfo{journal}{JHEP} \textbf{\bibinfo{volume}{11}}, \bibinfo{pages}{057}
  (\bibinfo{year}{2008}), \eprint{0810.0889}.

\bibitem[{\citenamefont{Avsar et~al.}(2009)\citenamefont{Avsar, Hatta, and
  Matsuo}}]{Avsar:2009yb}
\bibinfo{author}{\bibfnamefont{E.}~\bibnamefont{Avsar}},
  \bibinfo{author}{\bibfnamefont{Y.}~\bibnamefont{Hatta}}, \bibnamefont{and}
  \bibinfo{author}{\bibfnamefont{T.}~\bibnamefont{Matsuo}},
  \bibinfo{journal}{JHEP} \textbf{\bibinfo{volume}{06}}, \bibinfo{pages}{011}
  (\bibinfo{year}{2009}), \eprint{0903.4285}.

\bibitem[{\citenamefont{Agashe et~al.}(2008)\citenamefont{Agashe,
  Gopalakrishna, Han, Huang, and Soni}}]{Agashe:2008jb}
\bibinfo{author}{\bibfnamefont{K.}~\bibnamefont{Agashe}},
  \bibinfo{author}{\bibfnamefont{S.}~\bibnamefont{Gopalakrishna}},
  \bibinfo{author}{\bibfnamefont{T.}~\bibnamefont{Han}},
  \bibinfo{author}{\bibfnamefont{G.-Y.} \bibnamefont{Huang}}, \bibnamefont{and}
  \bibinfo{author}{\bibfnamefont{A.}~\bibnamefont{Soni}}
  (\bibinfo{year}{2008}), \eprint{0810.1497}.

\bibitem[{\citenamefont{Kuraev et~al.}(1977)\citenamefont{Kuraev, Lipatov, and
  Fadin}}]{Kuraev:1977fs}
\bibinfo{author}{\bibfnamefont{E.~A.} \bibnamefont{Kuraev}},
  \bibinfo{author}{\bibfnamefont{L.~N.} \bibnamefont{Lipatov}},
  \bibnamefont{and} \bibinfo{author}{\bibfnamefont{V.~S.} \bibnamefont{Fadin}},
  \bibinfo{journal}{Sov. Phys. JETP} \textbf{\bibinfo{volume}{45}},
  \bibinfo{pages}{199} (\bibinfo{year}{1977}).

\bibitem[{\citenamefont{Balitsky and Lipatov}(1978)}]{Balitsky:1978ic}
\bibinfo{author}{\bibfnamefont{I.~I.} \bibnamefont{Balitsky}} \bibnamefont{and}
  \bibinfo{author}{\bibfnamefont{L.~N.} \bibnamefont{Lipatov}},
  \bibinfo{journal}{Sov. J. Nucl. Phys.} \textbf{\bibinfo{volume}{28}},
  \bibinfo{pages}{822} (\bibinfo{year}{1978}).

\bibitem[{\citenamefont{Balitsky}(1996)}]{Balitsky:1995ub}
\bibinfo{author}{\bibfnamefont{I.}~\bibnamefont{Balitsky}},
  \bibinfo{journal}{Nucl. Phys.} \textbf{\bibinfo{volume}{B463}},
  \bibinfo{pages}{99} (\bibinfo{year}{1996}), \eprint{hep-ph/9509348}.

\bibitem[{\citenamefont{Kovchegov}(1999)}]{Kovchegov:1999yj}
\bibinfo{author}{\bibfnamefont{Y.~V.} \bibnamefont{Kovchegov}},
  \bibinfo{journal}{Phys. Rev.} \textbf{\bibinfo{volume}{D60}},
  \bibinfo{pages}{034008} (\bibinfo{year}{1999}), \eprint{hep-ph/9901281}.

\bibitem[{\citenamefont{Dasgupta et~al.}(2008)\citenamefont{Dasgupta, Magnea,
  and Salam}}]{Dasgupta:2007wa}
\bibinfo{author}{\bibfnamefont{M.}~\bibnamefont{Dasgupta}},
  \bibinfo{author}{\bibfnamefont{L.}~\bibnamefont{Magnea}}, \bibnamefont{and}
  \bibinfo{author}{\bibfnamefont{G.~P.} \bibnamefont{Salam}},
  \bibinfo{journal}{JHEP} \textbf{\bibinfo{volume}{02}}, \bibinfo{pages}{055}
  (\bibinfo{year}{2008}), \eprint{0712.3014}.

\bibitem[{\citenamefont{Butterworth et~al.}(2002)\citenamefont{Butterworth,
  Cox, and Forshaw}}]{Butterworth:2002tt}
\bibinfo{author}{\bibfnamefont{J.~M.} \bibnamefont{Butterworth}},
  \bibinfo{author}{\bibfnamefont{B.~E.} \bibnamefont{Cox}}, \bibnamefont{and}
  \bibinfo{author}{\bibfnamefont{J.~R.} \bibnamefont{Forshaw}},
  \bibinfo{journal}{Phys. Rev.} \textbf{\bibinfo{volume}{D65}},
  \bibinfo{pages}{096014} (\bibinfo{year}{2002}), \eprint{hep-ph/0201098}.

\bibitem[{\citenamefont{Almeida et~al.}(2009{\natexlab{a}})}]{Almeida:2008yp}
\bibinfo{author}{\bibfnamefont{L.~G.} \bibnamefont{Almeida}}
  \bibnamefont{et~al.}, \bibinfo{journal}{Phys. Rev.}
  \textbf{\bibinfo{volume}{D79}}, \bibinfo{pages}{074017}
  (\bibinfo{year}{2009}{\natexlab{a}}), \eprint{0807.0234}.

\bibitem[{\citenamefont{Butterworth et~al.}(2008)\citenamefont{Butterworth,
  Davison, Rubin, and Salam}}]{Butterworth:2008iy}
\bibinfo{author}{\bibfnamefont{J.~M.} \bibnamefont{Butterworth}},
  \bibinfo{author}{\bibfnamefont{A.~R.} \bibnamefont{Davison}},
  \bibinfo{author}{\bibfnamefont{M.}~\bibnamefont{Rubin}}, \bibnamefont{and}
  \bibinfo{author}{\bibfnamefont{G.~P.} \bibnamefont{Salam}},
  \bibinfo{journal}{Phys. Rev. Lett.} \textbf{\bibinfo{volume}{100}},
  \bibinfo{pages}{242001} (\bibinfo{year}{2008}), \eprint{0802.2470}.

\bibitem[{\citenamefont{Thaler and Wang}(2008)}]{Thaler:2008ju}
\bibinfo{author}{\bibfnamefont{J.}~\bibnamefont{Thaler}} \bibnamefont{and}
  \bibinfo{author}{\bibfnamefont{L.-T.} \bibnamefont{Wang}},
  \bibinfo{journal}{JHEP} \textbf{\bibinfo{volume}{07}}, \bibinfo{pages}{092}
  (\bibinfo{year}{2008}), \eprint{0806.0023}.

\bibitem[{\citenamefont{Kaplan et~al.}(2008)\citenamefont{Kaplan, Rehermann,
  Schwartz, and Tweedie}}]{Kaplan:2008ie}
\bibinfo{author}{\bibfnamefont{D.~E.} \bibnamefont{Kaplan}},
  \bibinfo{author}{\bibfnamefont{K.}~\bibnamefont{Rehermann}},
  \bibinfo{author}{\bibfnamefont{M.~D.} \bibnamefont{Schwartz}},
  \bibnamefont{and} \bibinfo{author}{\bibfnamefont{B.}~\bibnamefont{Tweedie}},
  \bibinfo{journal}{Phys. Rev. Lett.} \textbf{\bibinfo{volume}{101}},
  \bibinfo{pages}{142001} (\bibinfo{year}{2008}), \eprint{0806.0848}.

\bibitem[{\citenamefont{Almeida
  et~al.}(2009{\natexlab{b}})\citenamefont{Almeida, Lee, Perez, Sung, and
  Virzi}}]{Almeida:2008tp}
\bibinfo{author}{\bibfnamefont{L.~G.} \bibnamefont{Almeida}},
  \bibinfo{author}{\bibfnamefont{S.~J.} \bibnamefont{Lee}},
  \bibinfo{author}{\bibfnamefont{G.}~\bibnamefont{Perez}},
  \bibinfo{author}{\bibfnamefont{I.}~\bibnamefont{Sung}}, \bibnamefont{and}
  \bibinfo{author}{\bibfnamefont{J.}~\bibnamefont{Virzi}},
  \bibinfo{journal}{Phys. Rev.} \textbf{\bibinfo{volume}{D79}},
  \bibinfo{pages}{074012} (\bibinfo{year}{2009}{\natexlab{b}}),
  \eprint{0810.0934}.

\bibitem[{\citenamefont{Ellis et~al.}(2009)\citenamefont{Ellis, Vermilion, and
  Walsh}}]{Ellis:2009su}
\bibinfo{author}{\bibfnamefont{S.~D.} \bibnamefont{Ellis}},
  \bibinfo{author}{\bibfnamefont{C.~K.} \bibnamefont{Vermilion}},
  \bibnamefont{and} \bibinfo{author}{\bibfnamefont{J.~R.} \bibnamefont{Walsh}}
  (\bibinfo{year}{2009}), \eprint{0903.5081}.

\bibitem[{\citenamefont{Abbott et~al.}(1998)}]{Abbott:1998jb}
\bibinfo{author}{\bibfnamefont{B.}~\bibnamefont{Abbott}} \bibnamefont{et~al.}
  (\bibinfo{collaboration}{D0}), \bibinfo{journal}{Phys. Lett.}
  \textbf{\bibinfo{volume}{B440}}, \bibinfo{pages}{189} (\bibinfo{year}{1998}),
  \eprint{hep-ex/9809016}.

\bibitem[{\citenamefont{Abe et~al.}(1998)}]{Abe:1997ie}
\bibinfo{author}{\bibfnamefont{F.}~\bibnamefont{Abe}} \bibnamefont{et~al.}
  (\bibinfo{collaboration}{CDF}), \bibinfo{journal}{Phys. Rev. Lett.}
  \textbf{\bibinfo{volume}{80}}, \bibinfo{pages}{1156} (\bibinfo{year}{1998}).

\bibitem[{\citenamefont{Mueller and Tang}(1992)}]{Mueller:1992pe}
\bibinfo{author}{\bibfnamefont{A.~H.} \bibnamefont{Mueller}} \bibnamefont{and}
  \bibinfo{author}{\bibfnamefont{W.-K.} \bibnamefont{Tang}},
  \bibinfo{journal}{Phys. Lett.} \textbf{\bibinfo{volume}{B284}},
  \bibinfo{pages}{123} (\bibinfo{year}{1992}).

\bibitem[{\citenamefont{Del~Duca and Tang}(1993)}]{DelDuca:1993pq}
\bibinfo{author}{\bibfnamefont{V.}~\bibnamefont{Del~Duca}} \bibnamefont{and}
  \bibinfo{author}{\bibfnamefont{W.-K.} \bibnamefont{Tang}},
  \bibinfo{journal}{Phys. Lett.} \textbf{\bibinfo{volume}{B312}},
  \bibinfo{pages}{225} (\bibinfo{year}{1993}), \eprint{hep-ph/9304296}.

\bibitem[{\citenamefont{Cox et~al.}(1999)\citenamefont{Cox, Forshaw, and
  Lonnblad}}]{Cox:1999dw}
\bibinfo{author}{\bibfnamefont{B.}~\bibnamefont{Cox}},
  \bibinfo{author}{\bibfnamefont{J.~R.} \bibnamefont{Forshaw}},
  \bibnamefont{and} \bibinfo{author}{\bibfnamefont{L.}~\bibnamefont{Lonnblad}},
  \bibinfo{journal}{JHEP} \textbf{\bibinfo{volume}{10}}, \bibinfo{pages}{023}
  (\bibinfo{year}{1999}), \eprint{hep-ph/9908464}.

\bibitem[{\citenamefont{Enberg et~al.}(2002)\citenamefont{Enberg, Ingelman, and
  Motyka}}]{Enberg:2001ev}
\bibinfo{author}{\bibfnamefont{R.}~\bibnamefont{Enberg}},
  \bibinfo{author}{\bibfnamefont{G.}~\bibnamefont{Ingelman}}, \bibnamefont{and}
  \bibinfo{author}{\bibfnamefont{L.}~\bibnamefont{Motyka}},
  \bibinfo{journal}{Phys. Lett.} \textbf{\bibinfo{volume}{B524}},
  \bibinfo{pages}{273} (\bibinfo{year}{2002}), \eprint{hep-ph/0111090}.

\bibitem[{\citenamefont{Motyka et~al.}(2002)\citenamefont{Motyka, Martin, and
  Ryskin}}]{Motyka:2001zh}
\bibinfo{author}{\bibfnamefont{L.}~\bibnamefont{Motyka}},
  \bibinfo{author}{\bibfnamefont{A.~D.} \bibnamefont{Martin}},
  \bibnamefont{and} \bibinfo{author}{\bibfnamefont{M.~G.}
  \bibnamefont{Ryskin}}, \bibinfo{journal}{Phys. Lett.}
  \textbf{\bibinfo{volume}{B524}}, \bibinfo{pages}{107} (\bibinfo{year}{2002}),
  \eprint{hep-ph/0110273}.

\bibitem[{\citenamefont{Chevallier et~al.}(2009)\citenamefont{Chevallier,
  Kepka, Marquet, and Royon}}]{Chevallier:2009cu}
\bibinfo{author}{\bibfnamefont{F.}~\bibnamefont{Chevallier}},
  \bibinfo{author}{\bibfnamefont{O.}~\bibnamefont{Kepka}},
  \bibinfo{author}{\bibfnamefont{C.}~\bibnamefont{Marquet}}, \bibnamefont{and}
  \bibinfo{author}{\bibfnamefont{C.}~\bibnamefont{Royon}}
  (\bibinfo{year}{2009}), \eprint{0903.4598}.

\bibitem[{\citenamefont{Forshaw et~al.}(2005)\citenamefont{Forshaw, Kyrieleis,
  and Seymour}}]{Forshaw:2005sx}
\bibinfo{author}{\bibfnamefont{J.~R.} \bibnamefont{Forshaw}},
  \bibinfo{author}{\bibfnamefont{A.}~\bibnamefont{Kyrieleis}},
  \bibnamefont{and} \bibinfo{author}{\bibfnamefont{M.~H.}
  \bibnamefont{Seymour}}, \bibinfo{journal}{JHEP}
  \textbf{\bibinfo{volume}{06}}, \bibinfo{pages}{034} (\bibinfo{year}{2005}),
  \eprint{hep-ph/0502086}.

\bibitem[{\citenamefont{Bartels et~al.}(1995)}]{Bartels:1995rn}
\bibinfo{author}{\bibfnamefont{J.}~\bibnamefont{Bartels}} \bibnamefont{et~al.},
  \bibinfo{journal}{Phys. Lett.} \textbf{\bibinfo{volume}{B348}},
  \bibinfo{pages}{589} (\bibinfo{year}{1995}), \eprint{hep-ph/9501204}.

\bibitem[{\citenamefont{Gotsman et~al.}(1998)\citenamefont{Gotsman, Levin, and
  Maor}}]{Gotsman:1998mm}
\bibinfo{author}{\bibfnamefont{E.}~\bibnamefont{Gotsman}},
  \bibinfo{author}{\bibfnamefont{E.}~\bibnamefont{Levin}}, \bibnamefont{and}
  \bibinfo{author}{\bibfnamefont{U.}~\bibnamefont{Maor}},
  \bibinfo{journal}{Phys. Lett.} \textbf{\bibinfo{volume}{B438}},
  \bibinfo{pages}{229} (\bibinfo{year}{1998}), \eprint{hep-ph/9804404}.

\bibitem[{\citenamefont{Kaidalov et~al.}(2001)\citenamefont{Kaidalov, Khoze,
  Martin, and Ryskin}}]{Kaidalov:2001iz}
\bibinfo{author}{\bibfnamefont{A.~B.} \bibnamefont{Kaidalov}},
  \bibinfo{author}{\bibfnamefont{V.~A.} \bibnamefont{Khoze}},
  \bibinfo{author}{\bibfnamefont{A.~D.} \bibnamefont{Martin}},
  \bibnamefont{and} \bibinfo{author}{\bibfnamefont{M.~G.}
  \bibnamefont{Ryskin}}, \bibinfo{journal}{Eur. Phys. J.}
  \textbf{\bibinfo{volume}{C21}}, \bibinfo{pages}{521} (\bibinfo{year}{2001}),
  \eprint{hep-ph/0105145}.

\bibitem[{\citenamefont{Sung}(2009)}]{Sung:2009iq}
\bibinfo{author}{\bibfnamefont{I.}~\bibnamefont{Sung}} (\bibinfo{year}{2009}),
  \eprint{0908.3688}.

\end{thebibliography}

\end{document}